\documentclass{aa}  

\usepackage{placeins}
\usepackage{natbib}
\usepackage{amsmath}
\usepackage{amsfonts}
\usepackage{amssymb}
\usepackage{graphicx}
\usepackage[titletoc]{appendix}
\usepackage{color}
\usepackage{hyperref}
\usepackage{cleveref}
\usepackage[rightcaption]{sidecap}
\usepackage{subfigure}
\usepackage{comment}
\usepackage{soul}
\usepackage{cancel}
\usepackage{dcolumn}

\usepackage{array}
\usepackage{ctable}
\usepackage{multirow}
\usepackage{siunitx}
\usepackage{longtable}
\usepackage{tabularx}
\usepackage{booktabs}
\usepackage{float}

\graphicspath{{Graphics/}}

\def\be{\begin{equation}}
\def\ee{\end{equation}}
\def\bea{\begin{eqnarray}}
\def\eea{\end{eqnarray}}

\def\apjl{ApJ Lett.}
\def\apjs{ApJ Suppl. Ser.}

\def\aap{A\&A}

\definecolor{vividviolet}{rgb}{0.62, 0.0, 1.0}
\definecolor{amaranth}{rgb}{0.9, 0.17, 0.31}
\definecolor{palatinateblue}{rgb}{0.15, 0.23, 0.89}
\definecolor{brightpink}{rgb}{1.0, 0.0, 0.5}
\definecolor{cornflowerblue}{rgb}{0.39, 0.58, 0.93}
\definecolor{deepcarminepink}{rgb}{0.94, 0.19, 0.22}
\definecolor{radicalred}{rgb}{1.0, 0.21, 0.37}

\hypersetup{ linktoc=all,
    colorlinks, linkcolor={palatinateblue},
    citecolor={brightpink}, urlcolor={amaranth}
}

\DeclareUnicodeCharacter{2212}{\ensuremath{-}}

\begin{document} 

\title{No evidence for dynamical dark energy from the Combo correlation of gamma-ray bursts}

\author{Marco Muccino\inst{1,2,3,4}
\and
Massimo Della Valle\inst{5}
\and 
Luca Izzo\inst{6,7}
\and
Orlando Luongo\inst{1,2,8,9}}

\institute{Universit\`a di Camerino, Divisione di Fisica, Via Madonna delle carceri 9, 62032 Camerino, Italy.
\and
Al-Farabi Kazakh National University, Al-Farabi av. 71, 050040 Almaty, Kazakhstan.
\and
INAF, Catania Astrophysical Observatory, Via S.Sofia 79, 95123 Catania, Italy.
\and
ICRANet, Piazza della Repubblica 10,  65122 Pescara, Italy.
\and
INAF, Osservatorio Astronomico di Padova, Vicolo dell’Osservatorio 5, 35122 Padova, Italy
\and
INAF, Osservatorio Astronomico di Capodimonte, Salita Moiariello 16, 80131 Naples, Italy.
\and
DARK, Niels Bohr Institute, University of Copenhagen, Jagtvej 128, 2200 Copenhagen, Denmark.
\and
INAF, Osservatorio Astronomico di Brera, Milano, Italy.
\and
Department of Nanoscale Science and Engineering, University at Albany SUNY, Albany, NY 12222, USA.\\
\email{marco.muccino@unicam.it,massimo.dellavalle@inaf.it,luca.izzo@inaf.it,orlando.luongo@unicam.it}}
             
\date{}

\abstract
{Recently, the Dark Energy Spectroscopic Instrument (DESI) collaboration has presented results indicating that dark energy may exhibit dynamical behavior.}
{Calibrated gamma-ray burst (GRB) correlations can be employed to verify or reject a time-evolution of the dark energy (DE) equation of state, $\omega(z)$, up to redshifts $z\sim 9$.}
{We use the most updated catalog of GRBs fulfilling the Combo correlation and improve its calibration employing three catalogs of type Ia supernovae at redshifts $z\leq0.075$ and the B\'ezier interpolation of the Hubble rate, as an alternative to the cosmographic series that fails to be constraining at high redshifts. To test the evolution of $\omega(z)$, we adopt a model-independent, redshift-binned DE parametrization. In both the calibration and the DE reconstruction analyses the impact of the spatial curvature on the results is explored.}
{The calibrated Combo correlation yields a Hubble constant $H_0\sim70$~km/s/Mpc which alleviates the existing Hubble tension and is broadly consistent with current measurements, although the uncertainties prevent a high-precision measurement. Regarding the reconstruction of $\omega(z)$ of DE, spatially curved scenarios are disfavored and, despite the apparent ``phantom'' behavior at $z\lesssim0.55$ due to the limited statistics caused by the shortage of nearby events, at $z>0.55$ the analysis provides statistically robust evidence in favor of the cosmological constant scenario.}
{The Combo correlation alleviates the Hubble tension and shows no significant evidence in favor of dynamical DE. This suggests that GRBs, as distance indicators, are broadly consistent with the current cosmic distance ladder.}


\titlerunning{No evidence for dynamical dark energy from the Combo correlation of GRBs}

\authorrunning{Muccino et al.}
   
\maketitle

\section{Introduction}\label{sec:1}

The standard $\Lambda$CDM model is the simplest and most successful paradigm capable of fitting early- and late-time cosmological data \citep{Planck2018}. 
It attributes the late-time accelerated expansion of the universe, first evidenced by type Ia supernovae \citep[SNe Ia,][]{1998Natur.391...51P,Perlmutter1999,Riess1998,Schmidt1998} as driven by a form of \emph{dark energy} (DE) in the form of a cosmological constant $\Lambda$ with an equation of state (EoS) fixed at $\omega=-1$ at every redshift $z$, associated with vacuum energy \citep{2023CQGra..40j5004B}.

However, the advent of the \emph{precision cosmology era} \citep{2022JHEAp..34...49A} has revealed significant observational discrepancies, including tensions in the Hubble constant $H_0$ measurements \citep{Planck2018,2022ApJ...934L...7R} and, more notably, recent results based on baryonic acoustic oscillations (BAO) measurements from the Dark Energy Spectroscopic Instrument (DESI) favoring an evolving DE with redshift $z$ over a pure cosmological constant \citep{2025arXiv250314738D}. 

Focusing on the evolving DE, DESI-BAO data are in mild tension (at $2.3$~$\sigma$) with the flat $\Lambda$CDM model \citet{Planck2018} and show a preference towards the CPL \citep{Chevallier2001,Linder2003} dynamical DE parametrization $\omega(z) = \omega_0 + \omega_{\rm a} z/(1+z)$, with $w_0>−1$ and $w_{\rm a}<0$. 
However, DESI results are subject to several criticism, some of them even going so far as to question the reliability of some specific data points \citep[see, e.g.,][and references therein]{2024arXiv240408633C,2024arXiv241201740S,2025PhRvD.111b3512C,2025MNRAS.538..875E,2024Univ...11...10G,2025A&A...693A.187L,2025arXiv250210506S}.

In discerning between a possible \emph{revival} of dynamical DE and the cosmological constant paradigm, gamma-ray bursts (GRBs) may serve as crucial high-redshift tools, being detectable up to $z \approx 9$ \citep{Cucchiara2011,2012ApJ...749...68S}.
In establishing these sources as promising cosmic probes, several empirical correlations -- linking various GRB observables -- have been proposed \citep[see, e.g.,][and references therein]{Amati2002,Yonetoku2004,Ghirlanda2004,Schaefer2007,Dainotti2008,CapozzielloIzzo2008,Bernardini2012,Wei2014,Izzo2015}. 
All these attempts, once properly calibrated to alleviate the \emph{circularity problem} \citep[details in][]{2021Galax...9...77L}, have demonstrated that GRBs may provide meaningful cosmological constraints.

In this work, we use the most updated GRB sample from the Combo correlation \citep{Izzo2015,2021ApJ...908..181M} and further improve the calibration method introduced in \citet{Izzo2015} to overcome the circularity problem by utilizing three samples of SNe Ia up to redshifts $z\leq 0.075$ and resorting to the B\'ezier interpolation method \citep[see, e.g.,][]{orlando2} to bridge the calibration procedure from low to high redshifts, where cosmography fails.
Then, to study DE EoS, we adopt the model-independent, redshift-binned DE parametrization introduced in \citet{Daly2004}.
We consider two different redshift-binned analyses and, for each of them, we investigate the impact of the spatial curvature $\Omega_k$ by comparing the corresponding findings.
Our results are statistically robust, alleviate the existing Hubble tension, and favor a spatially flat universe with a DE in the form of a cosmological constant at $z>0.55$, whereas at $z\leq0.55$ results remain inconclusive. 
However, a certain interplay between $\Omega_k$ and dynamical DE cannot be excluded \emph{a priori}.

The paper is structured as follows.
In Sec.~\ref{sec:2}, we introduce the updated sample of Combo-GRBs. 
In Sec.~\ref{sec:3}, we work out the new calibration procedure and apply it to the updated Combo sample.
In Sec.~\ref{sec:4}, we test the possible evolution of DE with a piecewise formulation over GRB redshift intervals. 
In Sec.~\ref{sec:5}, we draw our conclusions.

\section{The updated Combo sample}\label{sec:2}

The Combo correlation (see \citealt{Izzo2015} and \citealt{2021ApJ...908..181M}, for details) is a hybrid prompt-afterglow emission correlation fulfilled by long GRBs and defined as
\begin{equation}
\label{eq:combo}
\log{L_0} = \alpha + \beta \log{E_{\rm p}} - \log{T}\ ,
\end{equation}
with intercept $\alpha$, slope $\beta$, and characterized by a dispersion parameter $\sigma$ that accounts for systematics and additional hidden uncertainties \citep{Dago2005}.

Eq.~\eqref{eq:combo} correlates the following observables:
\begin{itemize}
\item[-] the rest-frame peak energy $E_{\rm p}$ (in keV units), inferred from the spectrum of the GRB prompt emission, 
\item[-] the X-ray afterglow plateau luminosity $L_0 = 4\pi d_{\rm L}^2(z) F_0$ (in erg/s units), in the rest-frame $0.3$--$10$~keV band, related to the flux $F_0$ (in erg/cm$^2$/s units) and the source luminosity distance $d_{\rm L}(z)$, and
\item[-] the X-ray afterglow rest-frame effective duration of the plateau phase $T=\tau/|1+\gamma|$ (in s units), related to the rest-frame time $\tau$ (in s units) at which the late power-law decay phase with decay index $\gamma$ begins. 
\end{itemize}

\begin{figure}
\centering
\includegraphics[width=\hsize,clip]{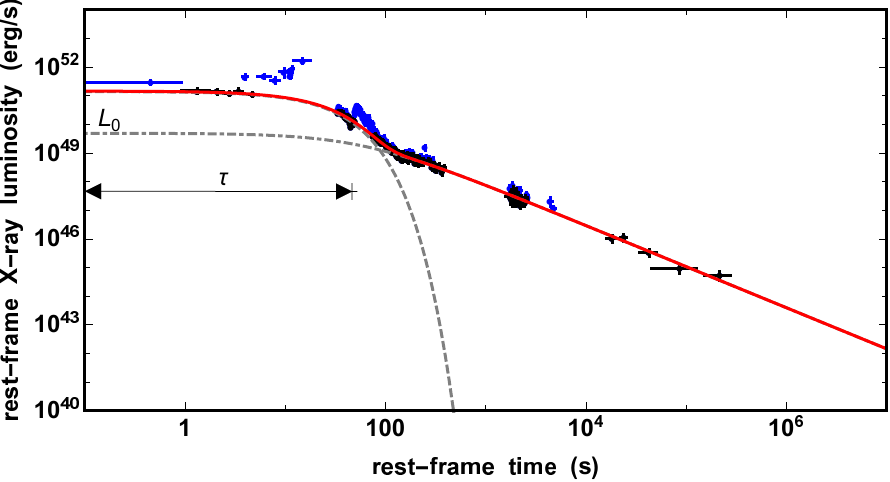}
\caption{Rest-frame $0.3$--$10$~keV LLC of GRB 060418A with the prompt + steep decay (dashed gray), the plateau + late decay (dot-dashed gray), and the total (solid red) best-fitting curves. The black dots are the data filtered by the flares (blue dots).}
\label{fig:MDV}
\end{figure}
In this work we expand the previous Combo sample composed of $174$ GRBs \citep{2021ApJ...908..181M}. To do so, below we summarize the adopted strategy.
\begin{itemize}
\item[1.] We consider GRBs detected from December $2004$ to December $2023$ with X-ray afterglow data from the \textit{Neil Gehrels Swift Observatory} \citep{Gehrels2004}.\footnote{\url{https://www.swift.ac.uk/burst_analyser/}} 
\item[2.] We retrieve the most updated GRB redshifts.\footnote{ \url{https://www.mpe.mpg.de/~jcg/grbgen.html}}
\item[3.] Updated $E_{\rm p}$ values come from \citet{2019ApJ...887...13F}, \citet{2021ApJ...908..181M}, \citet{2021ApJ...920..135X}, \citet{2022MNRAS.516.2575J}, and \citet{2024MNRAS.533..743W}. In case of multiple entries the latest value is considered and if the redshift does not match the value from step $2$, then $E_{\rm p}$ is duly rescaled. Finally, when no values are available in the above papers, those from GCN circulars are considered. 
\item[4.] Differently from \citet{Izzo2015} and \citet{2021ApJ...908..181M}, the X-ray afterglow steep decay phase is fit by the function $L_{\rm p}(t) = L_{\rm p0}\exp(-t/\tau_{\rm p})$, with normalization $L_{\rm p0}$ and decay time-scale $\tau_{\rm p}$ (see Fig.~\ref{fig:MDV}), which better describes this phase and refines the constraints on $L_0$, $\tau$, and $\alpha$. When available, prompt emission data from the \textit{Swift}-BAT instrument, transformed into the rest-frame $0.3$--$10$~keV band, are also included in the fit.
\item[5.] The observables $L_0$, $\tau$, and $\alpha$ are obtained by fitting the X-ray plateau and late-decay phase with the function $L(t)=L_0(1+t/\tau)^\gamma$ \citep[see][and Fig.~\ref{fig:MDV}]{RuffiniMuccino2014}.
\item[6.] The total profile $L_{\rm X}(t) = L_{\rm p}(t)+L(t)$ is then used to fit the rest-frame $0.3$--$10$~keV band, flare-filtered X-ray luminosity light curve (LLC) of the afterglow.\footnote{The LLC flare-filtering is performed iteratively: by fitting with $L_{\rm X}(t)$, at every iteration, data points with the largest positive residual are discarded, until a p-value $>0.3$ is obtained.}
\item[7.] LLCs with $\gamma>-1$ are excluded because they might have a change in slope beyond the temporal coverage of the data and/or be polluted by a late flaring activity.
\end{itemize}

\begin{figure}
\centering
\includegraphics[width=\hsize,clip]{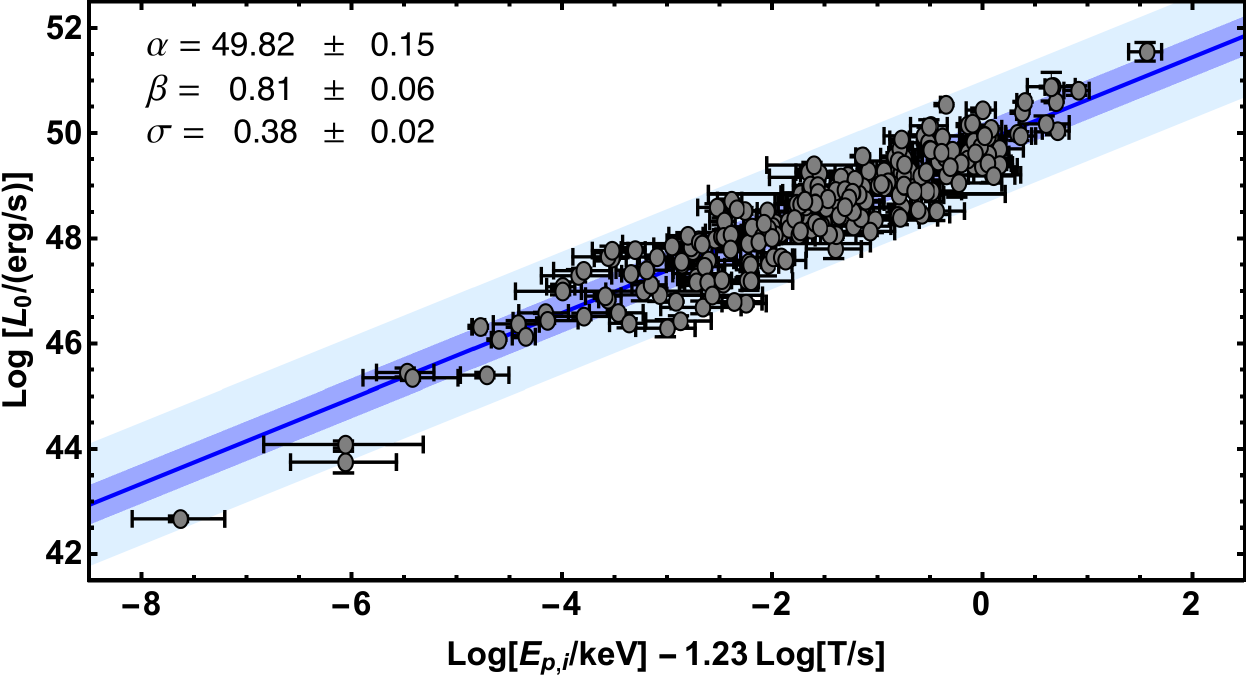}
\caption{The C244 sample (gray circles with errors) in the flat $\Lambda$CDM model with $H_0=67.36$~km/s/Mpc and $\Omega_m=0.3153$. The best-fit (solid blue line) and the $1$- and $3$-$\sigma$ bands (dark and light blue shaded regions, respectively) are also displayed.}
\label{fig:no}
\end{figure}

We obtain a sample of $N=244$ GRBs, hereby dubbed C244 (see Table~\ref{tab:no}). Fig.~\ref{fig:no} displays the updated Combo correlation, obtained by assuming the flat $\Lambda$CDM cosmology from \citet{Planck2018}. The best-fit parameters of the correlations are derived following the method of \citet{Dago2005}.

\section{The standardization method}\label{sec:3}

The proposed standardization method is based on the combination of three log-likelihood (LLH) functions.
\begin{itemize}
\item[(1)]{The first LLH, $\ln{\mathcal L_1}$, aims to determine $\beta$ and $\sigma$ from small sub-samples of GRBs lying almost at the same redshift \citep[see, e.g.,][]{Izzo2015,2021ApJ...908..181M};}
\item[(2)]{The second LLH, $\ln{\mathcal L_2}$, focuses on constraining the cosmic distance ladder by utilizing two \textit{calibrator} samples of SNe Ia from SBF and SH0ES and a \textit{cosmological} sample of SNe Ia in the redshift range $0.02\leq z\leq 0.075$ from the Pantheon catalog  \citep[see][]{2021A&A...647A..72K}. Because we want to extend it to large redshifts, where the cosmographic approach fails, we resort to the B\'ezier interpolation method \citep[see, e.g.,][]{orlando2}.}
\item[(3)]{The third LLH, $\ln{\mathcal L_3}$, uses the sub-samples identified in step (1) and the B\'ezier interpolation from step (2) to constrain the  intercept of the correlation.}
\end{itemize}
The use of distance moduli from SNe Ia in step (2) is restricted to redshifts $z\leq0.075$ where no GRBs of the C244 sample are present and various cosmological models degenerate, implying that differences in the distance moduli might be considered negligible. 

\subsection{The calibration of slope and extra-scatter parameters}\label{sec:3.1}

We identify $i=\{1,..,11\}$ sub-samples, each containing $j=\{1,..,N_i\}$ GRBs at approximately the same redshift $z_i$, thus providing well-determined correlations spanning at least two orders of magnitude in the Combo plane.

Since the $j$ GRBs of each sub-sample $i$ lie approximately at the same luminosity distance $d_{\rm L}(z_i)$, for each sub-sample we can fit the $i^{\rm th}$ correlation
\begin{equation}
\label{eq:combored}
\log{F_{0,ij}} = A_i + \beta_i \log{E_{{\rm p},ij}} - \log{T_{ij}}\,,
\end{equation}
where indexes $ij$ indicate the $j^{\rm th}$ GRB of the $i^{\rm th}$ sub-sample. 
For each sub-sample we introduce the rescaled intercepts $A_i=\alpha_i-\log{[4\pi d_{\rm L}^2(z_i)]}$ and indicate the intercepts $\alpha_i$, the slopes $\beta_i$, and the extra-scatters $\sigma_i$. These parameters can be determined for each sub-sample by defining the corresponding LLH functions in a cosmology-independent way
\begin{align}
\nonumber
\ln{\mathcal L_{1,i}} = &- \frac{1}{2} \sum_{j=1}^{N_i}\frac{(\log{F_{0,ij}}\! -\! A_i \!-\! \beta_i \log{E_{{\rm p},ij}}\!+\!\log{T_{ij}})^2}{\Sigma_{ij}^2}\\ 
\label{eq:deflike}
 &- \frac{1}{2} \sum_{j=1}^{N_i}\ln (2\pi \Sigma_{ij}^2)\,,
\end{align}
with $\Sigma_{ij}^2=\sigma_{\log{F_{0,ij}}}^2\!+\! \beta_i^2\sigma_{\log{E_{{\rm p},ij}}}^2 \!+\! \sigma_{\log{T_{ij}}}^2 \!+\! \sigma_i^2$.
The maximization of Eq.~\eqref{eq:deflike} can be performed in the parameter space
($\beta_i,\sigma_i$) by calculating $A_i$ and $\sigma_{A,i}$ from the condition $\partial\ln{\mathcal L_{1,i}}/\partial A_i=0$ \citep{2017A&A...598A.112D}
\begin{subequations}
\label{eq:res_int}
\begin{align}
\label{eq:deflikeA}
A_i &= \frac{\sum_{j=1}^{N_i} \left(\log{F_{0,ij}}\! - \beta_i \log{E_{{\rm p},ij}}\!+\!\log{T_{ij}}\right) \Sigma_{ij}^{-2}}
{\sum_{j=1}^{N_i} \Sigma_{ij}^{-2}}\,,\\
\sigma_{A,i} &= 
\frac{1}{\sqrt{\sum_{j=1}^{N_i} \Sigma_{ij}^{-2}}}\,.
\end{align}
\end{subequations}
The parameters $\beta_i$ and $\sigma_i$ of each sub-sample, determined by maximizing Eqs.~\eqref{eq:deflike}--\eqref{eq:res_int}, are listed in Table~\ref{tab:no3}.

\begin{table}
\centering
\tiny
\caption{Best-fit slope and extra-scatter for each sub-sample.}
\setlength{\tabcolsep}{1.3em}
\renewcommand{\arraystretch}{1.1}
\begin{tabular}{rcrcc}
\hline\hline
$\verb|#|$ & $z_i$           & $N_i$ & $\beta_i$         & $\sigma_i$ \\
\hline
$1$ & $0.54\pm0.01$	& $7$   & $0.69\pm0.45$     & $0.24\pm0.08$	\\
$2$ & $0.62\pm0.03$	& $7$   & $0.68\pm0.39$     & $0.46\pm0.14$	\\
$3$ & $1.12\pm0.05$	& $10$  & $0.90\pm0.22$	    & $0.14\pm0.06$	\\
$4$ & $1.25\pm0.04$	& $10$  & $0.94\pm0.42$   	& $0.36\pm0.10$	\\
$5$ & $1.48\pm0.04$	& $11$  & $0.88\pm0.53$	    & $0.22\pm0.09$	\\
$6$ & $1.74\pm0.06$	& $12$  & $1.15\pm0.33$	    & $0.26\pm0.09$	\\
$7$ & $2.05\pm0.05$	& $14$  & $0.92\pm0.35$	    & $0.40\pm0.10$ \\
$8$ & $2.27\pm0.07$	& $13$  & $0.63\pm0.26$	    & $0.28\pm0.07$ \\
$9$ & $2.56\pm0.15$	& $14$  & $0.78\pm0.17$	    & $0.20\pm0.09$	\\
$10$ & $2.98\pm0.09$	& $9$   & $0.76\pm0.19$	    & $0.18\pm0.07$ \\
$11$ & $3.59\pm0.09$	& $9$   & $0.97\pm0.32$	    & $0.21\pm0.08$ \\
\hline
\end{tabular}
\tablefoot{Columns list the progressive number, the mean redshift $z_i$, the number $N_i$ of GRBs, the slope $\beta_i$, and the dispersion $\sigma_i$.}
\label{tab:no3}
\end{table}

\begin{figure}
\centering
\hfill
\includegraphics[width=\hsize,clip]{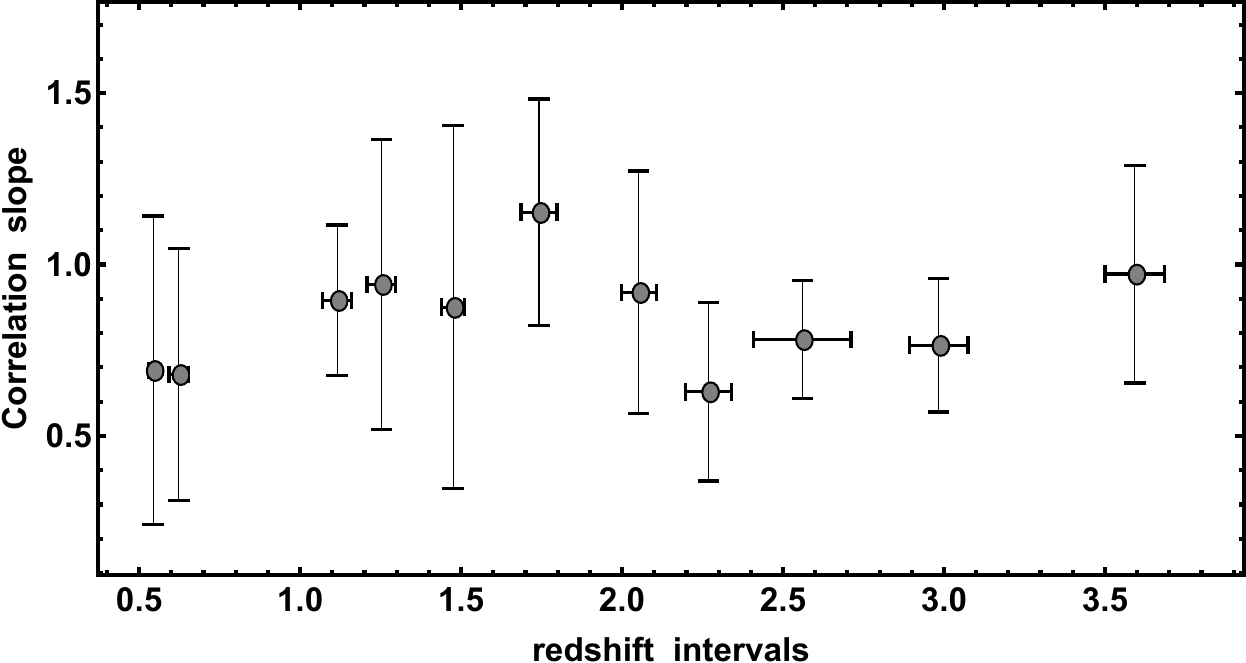}
\caption{The slopes $\beta_i$ from Table~\ref{tab:no3} of each sub-sample at $z_i$.}
\label{fig:no3}
\end{figure}

Focusing on the slope parameters $\beta_i$ shown in Fig.~\ref{fig:no3}, it is evident that the Combo correlation does not exhibit evolution effects over the $i$ redshift bins.
Motivated by this result, we calibrate the $i$ sub-samples assuming that they share the same intercept $\alpha$, slope $\beta$, and extra-scatter $\sigma$ (but different normalizations).
Thus, replacing $\alpha_i\rightarrow\alpha$ the various $A_i$ change only because of the differences in distances $d_{\rm L}(z_i)$; moreover, replacing $\beta_i\rightarrow\beta$ and $\sigma_i\rightarrow\sigma$ in Eq.~\eqref{eq:deflike}--\eqref{eq:res_int}, we determine $\beta$ and $\sigma$ by simultaneously fitting all the sub-samples and maximizing the total LLH 
\begin{equation}
\label{eq:LLH}
\ln{\mathcal L_1} = \sum_{i=1}^{11}\ln{\mathcal L_{1,i}}\,.
\end{equation}

\subsection{The calibration of the cosmic distance ladder via SNe Ia}\label{sec:3.2}

The choice of the calibrators is essential
in order to build a statistically significant sample of SNe Ia extending into the Hubble flow distances, given their relation with the $H_0$ value. 
To this aim, we hereby select: 
\begin{itemize}
\item the SBF calibrator sample, composed of $24$ SNe Ia with distance measurements evaluated through the SBF technique \citep{2021A&A...647A..72K},
\item the SH0ES calibrator sample, composed of $19$ SNe Ia with distances estimated using Cepheid variable stars \citep{riess2016}, and 
\item the cosmological sample, consisting of a subsample of $96$ SNe Ia from the Pantheon catalog \citep{pantheon} spanning a redshift range of $0.02\leq z \leq 0.075$ with the redshift cut $z < 0.02$ in order to mitigate the contamination from peculiar velocities \citep{riess2016}.
\end{itemize}

Differently from \citet{2021A&A...647A..72K}, where the SBF and the SH0ES samples were analyzed independently to compare the outcomes on the determination of $H_0$, here these samples are combined into a single one.
The rationale of this choice is evident after looking at the density distributions in Figs.~1--2 of \citet{2021A&A...647A..72K}: the SBF sample is characterized by early-type galaxies (dominated by old stellar populations) and low values of the ``color-stretch'' shape parameter $s$, whereas the SH0ES sample is composed of late-type spiral galaxies (being Cepheids relatively young stars) and high-valued $s$; the distribution of the cosmological sample lies between the two calibrator samples. 

For all the SNe Ia of the above samples, we utilize the best-fit apparent magnitudes at the light curve maximum in the B and V bands -- respectively $m_B$ and $m_V$ -- and $s$, as determined by using the ``max-model" method.\footnote{\url{https://github.com/nanditakhetan/SBF_SNeIa_H0} }

Once the light curve parameters are derived, the B-band peak apparent magnitude of each SN Ia can be modeled using the luminosity-stretch relation and the color correction accounting for the dust reddening in the host galaxy
\begin{equation}
\label{eq:Tripp}
m_B = P_0 + P_1 \left(s - 1\right) + R \,\Delta + \mu\,,
\end{equation}
where $P_0$ and $P_1$ give the luminosity-decline rate relation with $s$, $R$ is the extinction correction coefficient that correlates with the color $\Delta =m_B - m_V$, and $\mu$ is the distance modulus for the SN Ia host galaxy.

For the calibrator samples, the distance moduli in Eq.~\eqref{eq:Tripp} are those determined from SBF and Cepheids. 
For the cosmological sample, the distance moduli in Eq.~\eqref{eq:Tripp} are evaluated via a distance-redshift relation that differs from the usual cosmographic approach. This is motivated by the fact that cosmography fails to provide constraints at the high redshifts characterizing GRBs. Therefore, we resort to a model-independent interpolation of the Hubble rate $H(z)$
through a second order B\'ezier curve \citep{orlando2,2021MNRAS.503.4581L,2023MNRAS.518.2247L,2021MNRAS.501.3515M,2023MNRAS.523.4938M,2024JHEAp..42..178A,2024A&A...686A..30A,2024JCAP...12..055A,Alfano:2025fyq} 
\begin{equation}
\label{bezier1}
H(h,z) = h_\star z_{\rm m}^{-2} [h_0(z_{\rm m}-z)^2 + 2h_1 z(z_{\rm m}-z) +h_2 z^2]\,,
\end{equation}
with normalization $h_\star=100$~km/s/Mpc, maximum extrapolation redshift $z_{\rm m}=3.59\pm0.09$ given by the maximum redshift from Table~\ref{tab:no3}, and coefficients $h=\{h_0,h_1,h_2\}$. 
At $z=0$, Eq.~\eqref{bezier1} becomes $H_0 = h_\star h_0$, implying that $h_0$ coincides with the dimensionless Hubble constant.
From Eq.~\eqref{bezier1} we obtain the luminosity distance
\begin{equation}
\label{eq:da2}
d_{\rm L}(h,\Omega_k,z) = \frac{c\left(1+z\right)}{h_\star h_0\sqrt{\Omega_k}} \sinh \left[\int_0^z \frac{h_\star h_0\sqrt{\Omega_k} dz^\prime}{H(h_i,z^\prime)}\right]\,,
\end{equation}
and the corresponding distance modulus 
\begin{equation}
\label{eq:mu}
\mu(h,\Omega_k,z) = 5 \log [d_{\rm L}(h,\Omega_k,z)] + 25.
\end{equation}
that hold for any value of the curvature parameter $\Omega_k$.
 
From Eq.~\eqref{eq:Tripp}, the LLH functions for the calibrator (SBF~+~SH0ES) and the cosmological samples \citep[see details in][]{2021A&A...647A..72K} can be written as, respectively,
\begin{subequations}
\label{llcc}
\begin{align}
\ln \mathcal{L}_{cal} &= -\frac{1}{2}\sum_{i=1}^{N_{cal}} \left[ \frac{(m_B^i - m_B)^2}{\Sigma_{cal,i}^2} + \ln \left(2 \pi \Sigma_{cal,i}^2\right)\right]\,,\\
\ln \mathcal{L}_{cos} &= -\frac{1}{2}\sum_{j=1}^{N_{cos}} \left[ \frac{(m_B^j - m_B)^2}{\Sigma_{cos,j}^2} + \ln \left(2 \pi \Sigma_{cos,j}^2\right) \right]\,, 
\end{align}
\end{subequations}
and the full variances are, respectively,
\begin{subequations}
\label{sllcc}
\begin{align}
&\Sigma_{cal,i}^2 = (1-2R)\sigma_{m_B,i}^2\!+\! \sigma_{\mu,i}^2\!+\!P_1^2 \sigma_{s,i}^2\!+\!R^2 \sigma_{\Delta,i}^2\!+\!\sigma_{cal}^2,\\
&\Sigma_{cos,j}^2 = (1-2R)\sigma_{m_B,j}^2\!+\!P_1^2 \sigma_{s,j}^2 \!+\! R^2 \sigma_{\Delta,j}^2\!+\!\sigma_{cos}^2,
\end{align}
\end{subequations}
where $\sigma_{cal}$ (assumed to be the same for SBF and SH0ES) and $\sigma_{cos}$ are free parameters accounting for any extra dispersion observed in the measured distance moduli.

In conclusion, the total LLH function for the second step of the standardization method is given by
\begin{equation}
\label{LLH2}
    \ln \mathcal{L}_2 =  \ln \mathcal{L}_{cal} + \ln \mathcal{L}_{cos}\,.
\end{equation}

\subsection{The calibration of the intercept parameter}\label{sec:3.3}

We here resort Eqs.~\eqref{eq:res_int} defining $A_i$ and $\sigma_{A,i}$ in each sub-sample and the calibration of the cosmic ladder from Eqs.~\eqref{eq:Tripp}--\eqref{LLH2} to constrain the intercept of the Combo correlation via the LLH function
\begin{align}
\nonumber
\ln{\mathcal L_3} = &- \frac{1}{2} \sum_{i=1}^{11}\frac{[A_i - \alpha + \log{(4\pi)} + 2\log{d_{\rm L}(h,\Omega_k,z)}]^2}{\sigma_{A,i}^2}\\ 
\label{LLH3}
 &- \frac{1}{2} \sum_{i=1}^{11}\ln (2\pi \sigma_{A,i}^2)\,.
\end{align}

Unlike \citet{Izzo2015} and \citet{2021ApJ...908..181M}, where the intercept was calibrated using the distance moduli of only a small set of SNe Ia located at the redshift of the nearest GRBs of the Combo catalog, here the calibration utilizes three catalogs of SNe Ia, spanning a redshift range $z\leq0.075$, and eleven GRB sub-samples, spanning a redshift range $0.54\lesssim z \lesssim 3.59$.

\subsection{Results}\label{sec:3.4}

The best-fit parameters of the standardized Combo correlation can be inferred through an MCMC procedure by evaluating the posterior probability
\begin{equation}
\label{bayes}
\mathcal P(\theta|D,I) = \frac{\mathcal L(D|\theta,I) \mathcal{P}(\theta|I)}{\mathcal N(D|I)}\,,
\end{equation}
where, for given data $D$ and parameters $\theta$ with a set of assumptions $I$, $\mathcal L(D|\theta,I)$ is the likelihood function, $\mathcal{P}(\theta|I) $ is the prior probability, and $\mathcal N(D|I)$ is the normalization. 

Based on the definitions of Sec.~\ref{sec:3}, the likelihood function and the normalization are obtained by combining the LLH functions in Eqs.~\eqref{eq:deflike}, \eqref{LLH2}, and \eqref{LLH3}, i.e.,
\begin{equation}
\label{LLH_tot}
\ln\mathcal L = \ln\mathcal L_1 + \ln\mathcal L_2 + \ln\mathcal L_3\,.
\end{equation}
Regarding the prior probability $\mathcal{P}(\theta|I)$, for SN Ia calibration parameters $\theta_{SN}=\{P_0,P_1,R,\sigma_{cal},\sigma_{cos}\}$, Combo correlation parameters $\theta_C=\{\alpha,\beta,\sigma\}$, and the B\'ezier parameter $h_0$ (for which its uncertainty is dictated by the parameter $P_0$) we adopt uniform priors; regarding the B\'ezier parameters $h_1$ and $h_2$ and the curvature parameter $\Omega_k$, to get bounded constraints, we utilize Gaussian priors.
In view of these considerations, we have
\begin{equation}
\label{priors}
\mathcal{P}(\theta,I)\!\propto\!\left\{
\begin{array}{ll}
\!\displaystyle \exp\!\left[\!-\frac{(\theta_i-\bar\theta_i)^2}{2\sigma_{\theta,i}^2}\right], &\theta_i = \{h_1,h_2,\Omega_k\},\\
\!1, & \theta_i = \{\theta_{SN},\theta_{C},h_0\}.
\end{array}
\right.
\end{equation}
The mean values $\bar\theta_i$ and variances $\sigma_{\bar\theta,i}$ of the Gaussian priors are based on the combined MCMC results of the analyses from \citet{2025A&A...693A.187L}: $\bar h_1\pm\sigma_{\bar h_1}=1.01\pm0.24$, $\bar h_2\pm\sigma_{\bar h_2}=2.10\pm
0.36$, and $\bar \Omega_k\pm\sigma_{\bar \Omega_k}=-0.01\pm
0.22$.

We perform two MCMC analyses, using the Metropolis-Hastings algorithm, aiming at maximizing Eqs.~\eqref{bayes}--\eqref{priors}: the first labeled \emph{flat}, in which we fix $\Omega_k=0$, and the second labeled \emph{curved}, in which $\Omega_k$ is left free. 
The corresponding best-fit results are summarized in Table~\ref{tab:no5} and the MCMC posteriors are shown in Fig.~\ref{fig:no6a}.
To assess the statistically preferred model, we compute the Bayesian evidences and the difference $\Delta \ln B = \ln B_k-\ln B_0=0.82$. 
Although the Bayesian evidence shows a slight tendency toward the curved model, the difference is not statistically significant and therefore no preference can be established based on the modified Jeffreys’ scale \citep{2008ConPh..49...71T}.

\begin{table}
\setlength{\tabcolsep}{1.em}
\renewcommand{\arraystretch}{1.5}
\centering
\caption{Best-fit standardizations of C244 with SBF.}
\tiny
\begin{tabular}{lrr}
\hline\hline
Parameter   &  Flat  &  Curved \\
\hline
$P_0$        
&  $-19.22_{-0.08\,(0.12)}^{+0.06\,(0.11)}$
&  $-19.22_{-0.08\,(0.13)}^{+0.07\,(0.11)}$ \\
$P_1$  
& $-1.01_{-0.18\,(0.29)}^{+0.17\,(0.27)}$ 
& $-1.03_{-0.18\,(0.28)}^{+0.18\,(0.29)}$ \\
$R$ 
& $2.05_{-0.24\,(0.42)}^{+0.25\,(0.41)}$
& $2.01_{-0.23\,(0.39)}^{+0.29\,(0.45)}$ \\
$\sigma_{cal}$ 
& $0.25_{-0.05\,(0.07)}^{+0.05\,(0.10)}$ 
& $0.26_{-0.06\,(0.09)}^{+0.05\,(0.10)}$ \\
$\sigma_{cos}$ 
& $0.15_{-0.02\,(0.04)}^{+0.03\,(0.05)}$ 
& $0.15_{-0.02\,(0.04)}^{+0.03\,(0.05)}$ \\
$H_0$~(km/s/Mpc)  
& $70.44_{-2.27\,(3.75)}^{+2.38\,(3.97)}$ 
& $70.22_{-2.04\,(3.59)}^{+2.89\,(4.29)}$ \\
$h_1$ 
& $1.19_{-0.09\,(0.15)}^{+0.08\,(0.13)}$ 
& $1.17_{-0.09\,(0.14)}^{+0.08\,(0.14)}$ \\
$h_2$ 
& $2.39_{-0.21\,(0.34)}^{+0.21\,(0.36)}$
& $2.35_{-0.21\,(0.35)}^{+0.24\,(0.40)}$ \\
$\Omega_k$ 
& $0$
& $-0.26_{-0.24\,(0.39)}^{+0.25\,(0.42)}$ \\
$\alpha$ 
& $49.71_{-0.38\,(0.62)}^{+0.34\,(0.59)}$
& $49.71_{-0.33\,(0.53)}^{+0.38\,(0.58)}$ \\
$\beta$ 
& $0.92_{-0.14\,(0.23)}^{+0.14\,(0.24)}$ 
& $0.88_{-0.14\,(0.22)}^{+0.14\,(0.22)}$ \\
$\sigma$  
& $0.31_{-0.04\,(0.06)}^{+0.05\,(0.08)}$ 
& $0.31_{-0.04\,(0.06)}^{+0.04\,(0.08)}$ \\
\hline
$\ln \mathcal L$
& $-21.68$
& $-19.97$ \\
$\Delta\ln B$
& $0$
& $1.06$ \\
\hline
\end{tabular}
\tablefoot{For both $\Omega_k=0$ and $\Omega_k\neq0$ analyses, we report the Hubble constant value $H_0=h_\star h_0$ and, in the last two rows, LLH and difference of Bayesian evidence values.}
\label{tab:no5}
\end{table}

Table~\ref{tab:no5} also displays a
reduced $H_0$ tension between the values from Cepheid-calibrated SNe Ia, $H_0^{\rm R}=(73.04\pm1.04)$~km/s/Mpc \citep{2022ApJ...934L...7R}, and Cosmic Microwave Background (CMB), $H_0^{\rm P}=(67.36\pm0.54)$~km/s/Mpc \citep{Planck2018}. The agreement with $H_0^{\rm R}$ is at $\lesssim1$-sigma, whereas is at $\approx1.3$-sigma with $H_0^{\rm P}$.
Our values of $H_0$, determined from the joint SH0ES~+~SBF catalog, are consistent with the SBF-based, $H_0\sim70.5$~km/s/Mpc, and seem to indicate that SN Ia calibration is still biased and/or incomplete \citep[see][for details]{2021A&A...647A..72K}.

\begin{table}[t]
\setlength{\tabcolsep}{1.em}
\renewcommand{\arraystretch}{1.5}
\centering
\caption{Best-fit standardizations of C244 with IR--SBF.}
\tiny
\begin{tabular}{lrr}
\hline\hline
Parameter   &  Flat  &  Curved \\
\hline
$M$    
&  $-19.24_{-0.06\,(0.09)}^{+0.05\,(0.09)}$
&  $-19.23_{-0.06\,(0.09)}^{+0.05\,(0.09)}$ \\
$\alpha_{\rm S}$ 
& $0.19_{-0.02\,(0.04)}^{+0.02\,(0.03)}$ 
& $0.19_{-0.02\,(0.04)}^{+0.02\,(0.04)}$ \\
$\beta_{\rm S}$ 
& $2.72_{-0.31\,(0.49)}^{+0.31\,(0.52)}$
& $2.72_{-0.28\,(0.48)}^{+0.30\,(0.49)}$ \\
$\sigma_{cal}$ 
& $0.07_{-0.05\,(0.07)}^{+0.07\,(0.10)}$ 
& $0.09_{-0.07\,(0.09)}^{+0.05\,(0.09)}$ \\
$\sigma_{cos}$ 
& $0.13_{-0.02\,(0.04)}^{+0.02\,(0.04)}$ 
& $0.13_{-0.02\,(0.03)}^{+0.02\,(0.04)}$ \\
$H_0$~(km/s/Mpc)  
& $72.00_{-1.62\,(2.79)}^{+2.21\,(3.64)}$ 
& $72.34_{-1.92\,(3.18)}^{+1.87\,(3.22)}$ \\
$h_1$ 
& $1.17_{-0.08\,(0.13)}^{+0.09\,(0.15)}$ 
& $1.17_{-0.08\,(0.13)}^{+0.09\,(0.14)}$ \\
$h_2$ 
& $2.40_{-0.23\,(0.36)}^{+0.22\,(0.36)}$
& $2.40_{-0.24\,(0.38)}^{+0.19\,(0.34)}$ \\
$\Omega_k$ 
& $0$
& $-0.24_{-0.27\,(0.40)}^{+0.22\,(0.40)}$ \\
$\alpha$ 
& $49.69_{-0.38\,(0.59)}^{+0.37\,(0.58)}$
& $49.74_{-0.39\,(0.63)}^{+0.33\,(0.52)}$ \\
$\beta$ 
& $0.92_{-0.14\,(0.23)}^{+0.14\,(0.23)}$ 
& $0.89_{-0.15\,(0.23)}^{+0.13\,(0.22)}$ \\
$\sigma$  
& $0.31_{-0.04\,(0.06)}^{+0.05\,(0.09)}$ 
& $0.31_{-0.04\,(0.06)}^{+0.05\,(0.08)}$ \\
\hline
$\ln \mathcal L$
& $7.26$
& $9.20$ \\
$\Delta\ln B$
& $0$
& $0.83$ \\
\hline
\end{tabular}
\tablefoot{Same as in Table~\ref{tab:no5}. Details can be found in Appendix~\ref{appendix3}.}
\label{tab:no5b}
\end{table}

To further investigate the influence of the SBF technique in calibrating SNe Ia, we now perform a comparative analysis that involves as a calibrator sample a set of SNe Ia with infrared SBF (IR--SBF) distance measurements \citep{Garnavich:2022hef,Jensen:2021ooi}, for which only the SALT2 model parameters obtained from \texttt{SNANA} software package are available \citep{2022ApJ...938..113S}. Therefore, in combining this sample with the SNe of the SH0ES calibrator sample and the cosmological sample with $0.02\leq z \leq 0.075$, we have to adopt the same parametrization. The corresponding entries are retrieved from the online Pantheon+ catalog.\footnote{\url{https://github.com/PantheonPlusSH0ES/DataRelease}} The calibration procedure of the cosmic ladder obtained via this methodology is fully detailed in Appendix~\ref{appendix3}. The results, summarized in Table~\ref{tab:no5b} and shown in Fig.~\ref{fig:no6b}, still favor a spatially flat universe and provide similar values of the Combo correlation parameters. However the Hubble constant value, $H_0\sim72.00$~km/s/Mpc, is still lower than (but more consistent with) $H_0^{\rm R}$, whereas the tension with $H_0^{\rm P}$ is now more pronounced ($\approx2.6$-sigma).
\begin{figure}
\centering
{\includegraphics[width=0.95\hsize,clip]{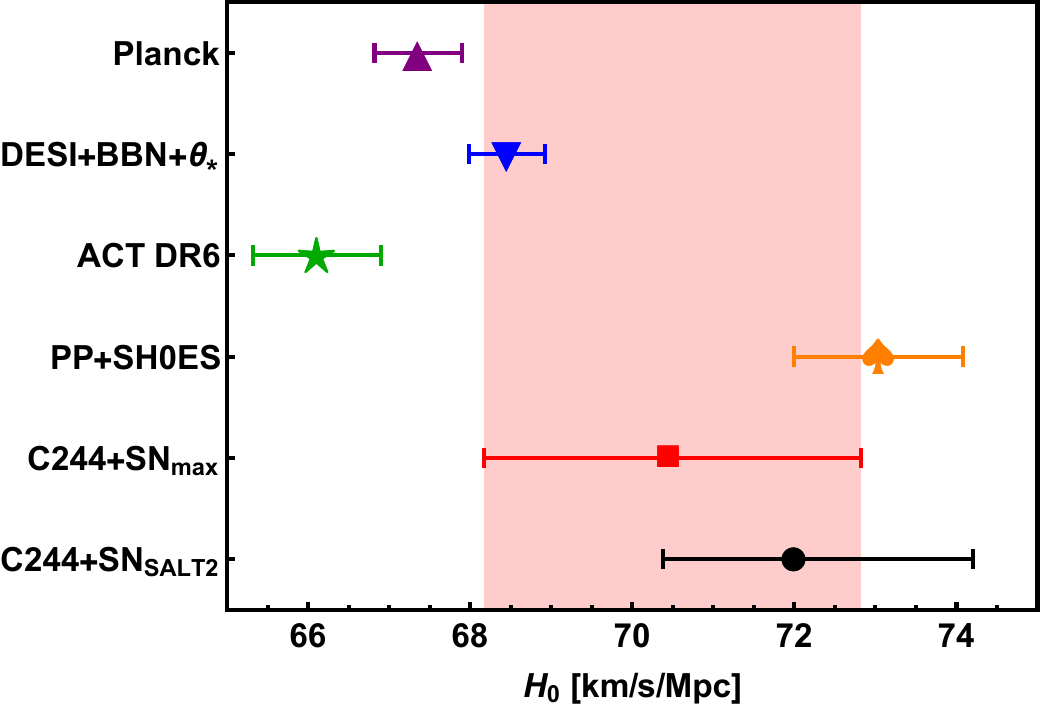}}
\caption{Comparison of $H_0$ constraints obtained from Planck (purple triangle), DESI+BBN+$\theta_\star$ (blue reversed triangle), ATC DR6 (green star), PP+SH0ES (orange spades), C244+SN$_{\rm max}$ (red square), and C244+SN$_{\rm SALT2}$ (black circle). The $1$-sigma constraint from C244+SN$_{\rm max}$ is marked with a light red shaded area to highlight the reduced tension between $H_0^{\rm R}$ (Planck) and $H_0^{\rm P}$ (PP+SH0ES). See text for details.}
\label{fig:comp}
\end{figure}

Fig.~\ref{fig:comp} portrays the two estimates of $H_0$ obtained by calibrating the C244 sample with $\Omega_k=0$ and using: a) SNe Ia with the ``max-model'' parametrization (C244+SN$_{\rm max}$) and b) SNe Ia with the SALT2 parametrization (C244+SN$_{\rm SALT2}$). These values are compared with $H_0^{\rm P}$ got from the CMB \citep{Planck2018}, the value obtained by combining DESI with the constraints of Big Bang Nucleosynthesis (BBN) and the acoustic angular scale $\theta_\star$ \citep{2025arXiv250314738D}, the measurement from the Data Release 6 maps made from Atacama Cosmology Telescope (ACT DR6, \citealt{AtacamaCosmologyTelescope:2025blo}), and the estimate from SNe Ia from the Pantheon+ catalog and SH0ES (PP+SH0ES), $H_0^{\rm R}$ \citep{2022ApJ...934L...7R}. 
In view of the above comparison, we select the calibration based on the use of the ``max-model'' method for the SBF sample, since this technique has the advantage of reducing the tension between $H_0^{\rm R}$ and $H_0^{\rm P}$.

\begin{figure*}
\centering
{\includegraphics[width=0.49\hsize,clip]{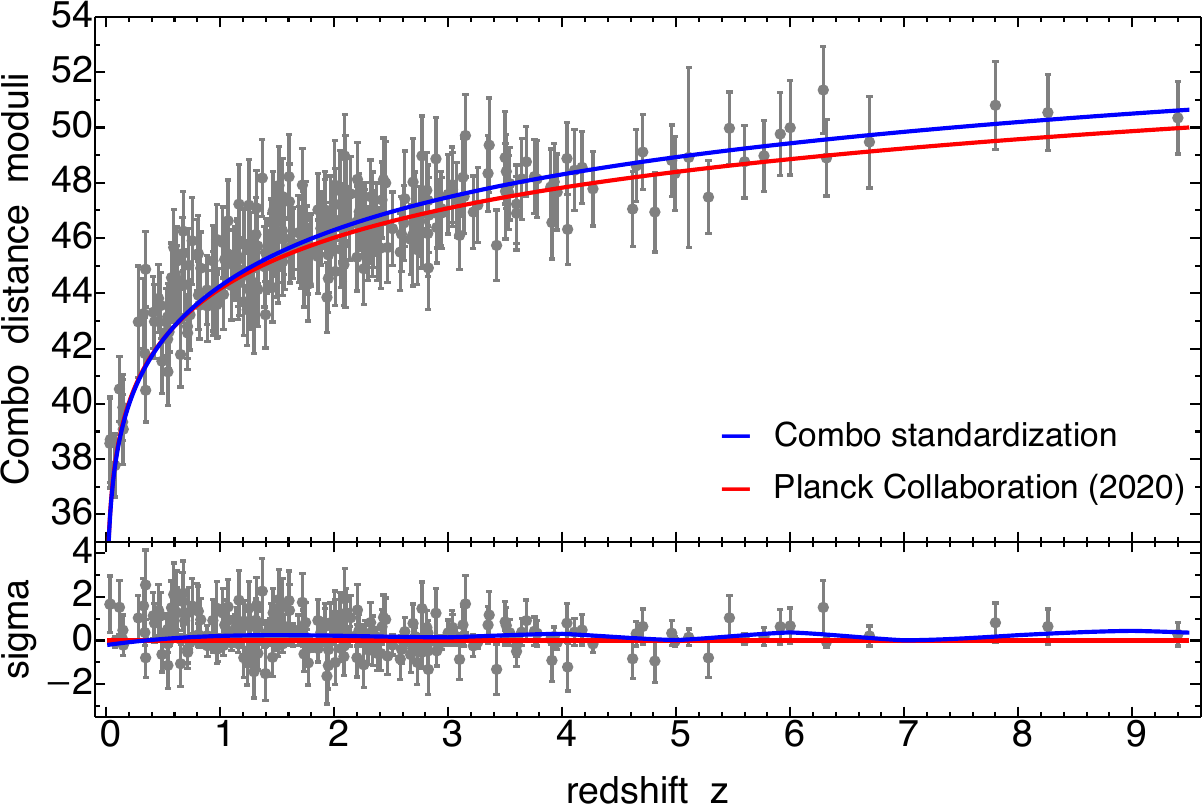}
\hfill
\includegraphics[width=0.49\hsize,clip]{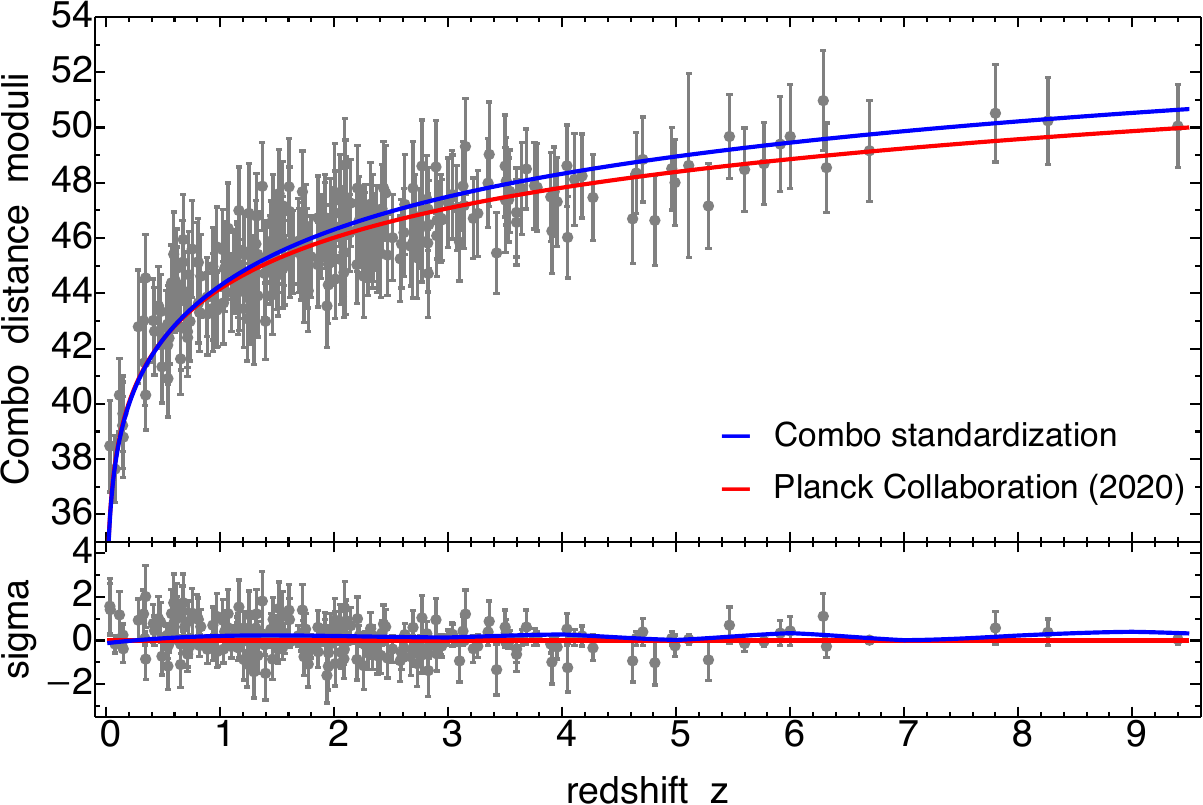}}
\caption{Standardized C244 distance moduli (gray data with errors) compared with standardized best-fit curves (solid blue) and the \citet{Planck2018} $\Lambda$CDM model (solid red) for $\Omega_k=0$ (left panel) and $\Omega_k\neq0$ (right panel).}
\label{fig:no5}
\end{figure*}

Combining Eqs.~\eqref{eq:combo} and \eqref{eq:mu} and using the definition $L_0 = 4\pi d_{\rm L}^2(z) F_0$, we define the C244 distance moduli as
\begin{subequations}
\label{eq:no3}
\begin{align}
\mu_{\rm GRB} &= \mu_0 + \frac{5}{2} \left(\alpha + \beta \log{E_{\rm p}} - \log{T} -\log{F_0} \right)\,,\\
\sigma_{\mu_{\rm GRB}} &= \frac{5}{2} \sqrt{\Sigma^2 + \sigma_\alpha^2 + (\log{E_{\rm p}})^2 \sigma_\beta^2 + 2 \log{E_{\rm p}}\sigma_{\alpha\beta}}\,,
\end{align}
\end{subequations}
where the quantity $\mu_0=-100.20$ accounts for all the constants and the conversion factor from Mpc to cm, the correlation error is $\Sigma^2=\sigma_{\log{F_0}}^2\!+\! \beta^2\sigma_{\log{E_{\rm p}}}^2 + \sigma_{\log T}^2 + \sigma^2$, and $\sigma_{\alpha\beta}$ is the covariance term between $\alpha$ and $\beta$.
Using the calibrated parameters in Table~\ref{tab:no5}, from Eqs.~\eqref{eq:no3} we obtain the distance moduli 
portrayed in Fig.~\ref{fig:no5}, where they are compared with Eq.~\eqref{eq:mu} for the best-fit parameters from Table~\ref{tab:no5}, obtained from the Combo standardization (see blue curve), and the $\Lambda$CDM distance moduli (see red curve) obtained from Eqs.~\eqref{eq:da2}--\eqref{eq:mu} using $H(z)=H_0\sqrt{\Omega_m(1+z)^3+1-\Omega_m}$, $H_0=67.36$~km/s/Mpc, $\Omega_k=0$, and $\Omega_m=0.3153$ \citep{Planck2018}.

Table~\ref{tab:no5} indicates no statistical preference for \emph{flat} and \emph{curved} models, in Table~\ref{tab:no} we list the distance moduli derived from both calibrations, respectively labeled as $\mu_{\rm GRB}(\Omega_k=0)$ and $\mu_{\rm GRB}(\Omega_k\neq0)$.

\section{Testing the evolution of dark energy}\label{sec:4}

To test DE scenarios with GRBs, their calibrated distance moduli $\mu_{\rm GRB}$ from Eq.~\eqref{eq:no3} are compared to the theoretical ones in Eq.~\eqref{eq:mu}, where the assumed cosmological model enters through the choice of $d_{\rm L}(z)$.
The corresponding most generic expression is given by Eq.~\eqref{eq:da2} where, accounting for pressureless matter, curvature, and any DE component -- with density parameters $\Omega_m$, $\Omega_k$, $\Omega_{\rm DE}=1-\Omega_m-\Omega_k$, respectively -- the Hubble rate takes the form 
\begin{equation}
\label{eq:no5}
H(z) = H_0 \sqrt{\Omega_m(1+z)^3 + \Omega_k(1+z)^2 + \Omega_{\rm DE} f(z)}\,,
\end{equation}
where $f(z)$ parametrizes the evolution of DE with $z$ and for the $\Lambda$CDM model $f(z)\equiv1$.

\subsection{Dynamical dark energy from DESI observations}\label{sec:4.1}

The results from the first two data releases of DESI \citep{2025arXiv250314738D} can be summarized as follows.
\begin{itemize}
\item[-] BAO data are consistent with a flat $\Lambda$CDM model, but the inferred parameters are in mild tension ($2.3$~$\sigma$) with the results of \citet{Planck2018}. 
\item[-] A CPL dynamical DE model with $w_0>−1$ and $w_{\rm a}<0$ is preferred over $\Lambda$CDM at $3.1$~$\sigma$ for DESI-BAO + CMB data and, including SNe Ia, at $2.8$--$4.2$~$\sigma$ depending on the considered SN sample. 
\item[-] Non-parametric methods consistently favor dynamical DE at redshifts $z\lesssim 0.3$ \citep{2025arXiv250314743L}.
\end{itemize}

However, these results raise several criticisms, some of them even going so far as to question DESI dataset itself.
\begin{itemize}
\item[-] The evidence for dynamical DE appears to be sensitive to the choice of SN data sets combined with DESI-BAO data, raising concerns about potential systematics in SN measurements influencing the results  \citep{2025MNRAS.538..875E}.
\item[-] Dynamical DE preference over the cosmological constant scenario may result from breaking the degeneracy between $H_0$ and the comoving sound horizon at the baryon drag epoch $r_d$ affecting the DESI-BAO data, as evidenced when combining them with observations like CMB or cosmic chronometers \citep{2025PhRvD.111b3512C}.
\item[-] Specific DE parametrizations show reduced tension with the $\Lambda$CDM model, indicating that the dynamical behavior may not be robust \citep{2025PhRvD.111b3512C,2024Univ...11...10G}. 
Conversely, model-independent approaches, like symbolic regression \citep{2025arXiv250210506S} or parametric interpolations \citep{2025A&A...693A.187L}, suggest that the $\Lambda$CDM model is still the best-suited one.
\item[-] Certain measurements within the DESI dataset have been identified as potential outliers that once excluded from analyses, reduce the tension with the $\Lambda$CDM model  \citep{2024arXiv240408633C,2024arXiv241201740S}.
\end{itemize}

In view of the above, it is therefore interesting to assess whether GRBs can provide valuable information in favor of or in contrast with the dynamical nature of DE.

\subsection{Redshift-binned parametrization of dark energy}\label{sec:4.2}

To assess the behavior of DE in a model-independent way, we consider in Eq.~\eqref{eq:no5} an $f(z)$ that makes no assumptions on the nature of DE \citep{2015arXiv151207076L}. Specifically, we employ the following redshift-binned DE parametrization \citep{Daly2004,King2014} 
\begin{equation}
\label{eq:fzbinned}
f(z)=(1+z)^{3(1+\tilde{\omega}_n)}\prod_{i=0}^{n-1}\left(1+\bar z_i\right)^{3(\tilde{\omega}_i-\tilde{\omega}_{i+1})},
\end{equation}
utilizing $n$ correlated DE barotropic indices $\tilde{\omega}_i$ -- each for the corresponding redshift bin defined by the maximum redshift $\bar z_i$ -- with covariance matrix given by
\begin{equation}
\textbf{C}=\langle {\bf \tilde{\omega}}{\bf \tilde{\omega}}^{\rm T}\rangle-\langle{\bf \tilde{\omega}}\rangle\langle{\bf \tilde{\omega}}^{\rm T}\rangle\ ,
\end{equation}
Following \citet{Huterer2005}, to decorrelate the parameters $\tilde{\omega}_i$ we need to: 
\begin{itemize}
\item[(a)] get the diagonal Fisher matrix ${\bf \Lambda}$ through an orthogonal matrix rotation \textbf{O}, i.e., ${\bf C}^{-1}={\bf O}^{\rm T}{\bf \Lambda} {\bf O}$,
\item[(b)] obtain the weight matrix $\textbf{W}=\textbf{O}^{\rm T}{\bf \Lambda}^{1/2}\textbf{O}$, normalized in such a way that its row elements sum to unity, and 
\item[(c)] decorrelate the parameters via ${\bf \omega}=\textbf{W}{\bf \tilde{\omega}}$.
\end{itemize}

\subsection{Results}\label{sec:4.3}

We conduct two analyses, based on two binning choices. In both, we fix the first bin, like in \citet{2021ApJ...908..181M}, to be $z\leq \bar z_1=0.55$ because a) it encloses the redshift range where the transition from a matter-dominated universe to a DE-dominated one begins and b) to guarantee that the presence of a sufficient number of low-$z$ GRBs.
Then, depending on the remaining bins we distinguish between:
\begin{itemize}
\item[-] {\bf Binning 1 (B1)}, which provides a parametrization of DE EoS based on a total of $5$ bins with intervals similar to those from \citet{2021ApJ...908..181M}, and 
\item[-] {\bf Binning 2 (B2)}, an averaged parametrization based on only two bins, with $z\leq \bar z_1$ or $z>\bar z_1$. 
\end{itemize}
Both Binnings $1$ and $2$ are split into two sub-analyses depending on whether curvature $\Omega_k$ is included or not. 
Thus, based on the results of Table~\ref{tab:no5} and the binning, we use the following GRB distance moduli and priors:
\begin{itemize}
\item[-] for $\Omega_k=0$, we use the distance moduli $\mu_{\rm GRB}(\Omega_k=0)$ from Table~\ref{tab:no} and the following Gaussian prior
\begin{equation}
\nonumber
\bar H_0\pm \sigma_{\bar H_0}=(70.44 \pm 2.33)~{\rm km/s/Mpc}\,;
\end{equation}
\item[-] for $\Omega_k\neq0$, we use the distance moduli $\mu_{\rm GRB}(\Omega_k\neq0)$ from Table~\ref{tab:no} and the following Gaussian priors
\begin{align}
\nonumber
&\bar H_0\pm \sigma_{\bar H_0}=(70.22 \pm 2.47)~{\rm km/s/Mpc}\,,\\
\nonumber
&\bar \Omega_k\pm \sigma_{\bar \Omega_k}=-0.26 \pm 0.25\,;
\end{align}
\item[-] for all the analyses, we impose uniform priors on the other parameters
\begin{equation}
\nonumber
\Omega_m\in[0,1]\,,\qquad \omega_i\in[-10,1]\,.
\end{equation}
\end{itemize}

\begin{table*}
\setlength{\tabcolsep}{.55em}
\renewcommand{\arraystretch}{1.4}
\centering
\tiny
\caption{Best-fit parameters from the redshift-binned DE parametrization, for B1 and B2.}
\begin{tabular}{c|cccccccccc}
\hline\hline
                          & $H_0$       &
$\Omega_k$                & $\Omega_m$  &
$\omega_1$                & $\omega_2$  &
$\omega_3$                & $\omega_4$  & 
$\omega_5$                & $\ln{\mathcal L_f}$ &
$\Delta\ln B$ \\
& (km/s/Mpc) & & & & & & & & & \\
\hline
B1                        & $69.37_{-3.30}^{+4.17}$ & 
--                        & $0.27_{-0.05}^{+0.06}$  &
$-4.60^{+1.09}_{-1.09}$   & $-2.54^{+1.18}_{-1.18}$ &
$-0.73^{+1.46}_{-1.46}$   & $-2.96^{+1.44}_{-1.44}$ &
$-0.46^{+1.03}_{-1.03}$   & $-379.93$               &
$6.99$\\ 
                          & $69.05_{-3.48}^{+4.05}$ & 
$-0.15_{-0.29}^{+0.20}$   & $0.51_{-0.14}^{+0.15}$  &
$-4.31^{+1.28}_{-1.28}$   & $-2.69^{+1.30}_{-1.30}$ &
$-1.36^{+1.28}_{-1.28}$   & $-2.09^{+1.50}_{-1.50}$ &
$-0.12^{+1.25}_{-1.25}$   & $-390.57$               &
$17.63$\\
\cline{5-9}
&&&                       & $\bar z_1=0.55$  & 
$\bar z_2=1.30$           & $\bar z_3=2.15$  & 
$\bar z_4=3.00$           & $\bar z_5=9.50$  & &\\
&&&                       & $n_1=22$         & 
$n_2=58$                  & $n_3=68$         & 
$n_4=45$                  & $n_5=51$         & &\\
\hline
B2                        & $69.91_{-4.12}^{+3.42}$ &
--                        & $0.28_{-0.05}^{+0.07}$  & $-3.97^{+1.53}_{-1.53}$   & $-0.85^{+1.28}_{-1.28}$ &  
&  &  & $-381.19$ & $0$ \\
                          & $69.58_{-3.86}^{+3.71}$ &
$-0.21_{-0.23}^{+0.23}$   & $0.50_{-0.14}^{+0.14}$  &
$-3.71^{+1.84}_{-1.84}$   & $-1.28^{+1.29}_{-1.29}$ &  
&  &  & $-391.22$ & $10.03$ \\ 
\cline{5-6}
&&&                       & $\bar z_1=0.55$  & 
$\bar z_2=9.50$           &  &  &  & & \\
&&&                       & $n_1=22$         & 
$n_2=222$                 &  &  &  & & \\
\hline\hline
\end{tabular}
\tablefoot{Column lists, respectively, the binning, the obtained $H_0$, $\Omega_k$ and $\Omega_m$, the DE parameters $\omega_i$, and the corresponding LLH and Bayes evidence values. Below each $\omega_i$ value we report the maximum redshift $\bar z_i$ and the number of sources $n_i$ within each bin.}
\label{tab:no6}
\end{table*}

The best-fit values are inferred via MCMC procedures by evaluating the posterior probability, as per Eq.~\eqref{bayes}, via priors defined as per Eq.~\eqref{priors}, and the LLH function
\begin{equation}
\label{eq:LLHf}
\ln{\mathcal L_f} = - \frac{1}{2} \sum_{i=1}^{N}\!\left\{\!\frac{\left[\mu_{{\rm GRB},i} - \mu_f(z_i)\right]^2}{\sigma_{\mu_{{\rm GRB},i}}^2}\!+\! \ln (2\pi \sigma_{\mu_{{\rm GRB},i}}^2)\!\right\},
\end{equation}
where $\mu_f$ are the theoretical distance moduli of Eq.~\eqref{eq:da2}--\eqref{eq:mu}, with the positions given by Eqs.~\eqref{eq:no5}--\eqref{eq:fzbinned}, and the GRB distance moduli with errors, $\mu_{{\rm GRB},i}\pm\sigma_{\mu_{{\rm GRB},i}}$, for both \emph{flat} and \emph{curved} calibrations, are listed in Table~\ref{tab:no}. 

\begin{figure*}
\centering
\includegraphics[width=0.94\hsize,clip]{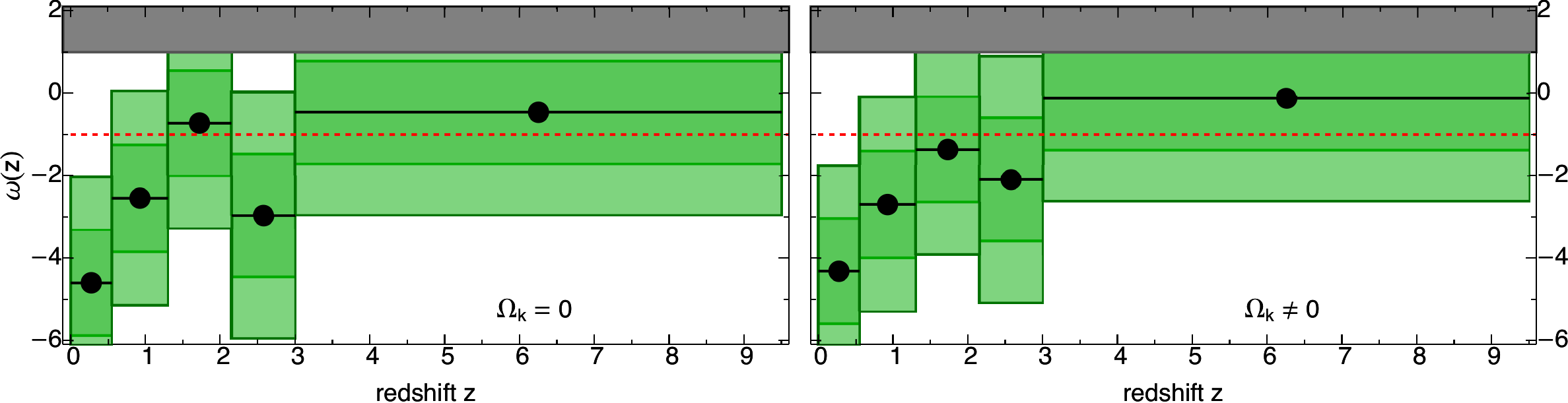}
\includegraphics[width=0.94\hsize,clip]{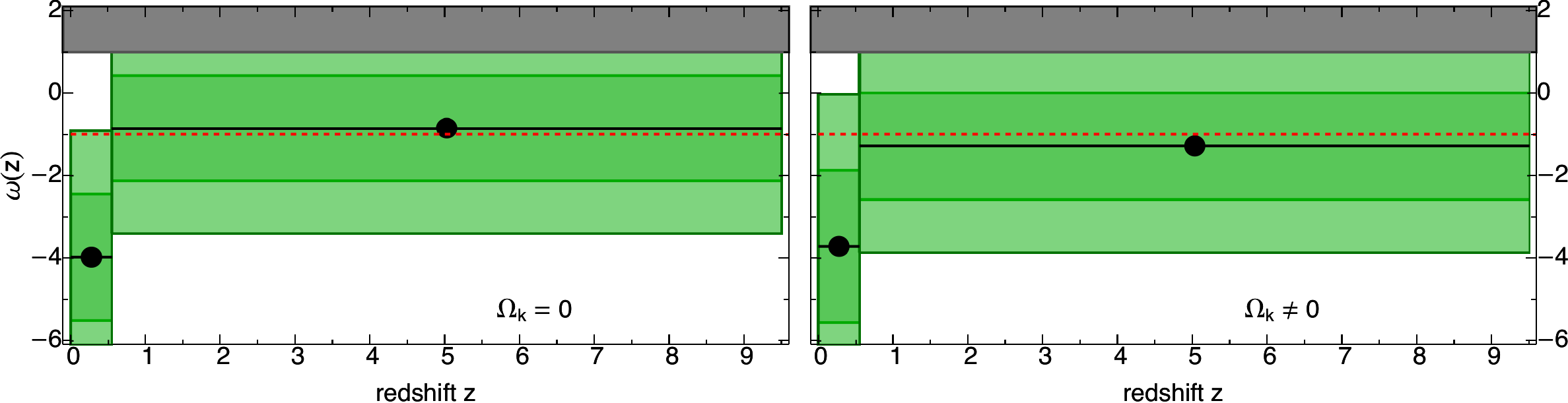}
\caption{B1 (top) and B2 (bottom) DE reconstructed $\omega_i$ ($1$ and $2\sigma$ in dark and light green, respectively), for flat (left) and curved (right) cases, compared to $\omega=-1$ of the flat $\Lambda$CDM model (dashed red lines). The dark gray regions exclude unphysical EoS.}
\label{fig:no6}
\end{figure*}

The results, with the DE parameters already decorrelated according to the procedure outlined in Sec.~\ref{sec:4.2}, are listed in Table~\ref{tab:no6} and portrayed in Fig.~\ref{fig:no6}, where unphysical, beyond-stiff EoS regions with $\omega>1$ are excluded.
Below, we summarize them, focusing on specific points and performing comparisons with previous results.
\begin{itemize}
\item[1)] For $\Omega_k=0$, the values of $\Omega_m$ are consistent with those from both the \emph{Pantheon}+ catalog of SNe Ia \citep{2022ApJ...938..110B} and CMB \citep{Planck2018}.
\item[2)] For $\Omega_k\neq0$, we obtain values of $\Omega_m$ inconsistent (at $1\sigma$) with both SNe Ia and CMB. However, this is not a surprising outcome when involving GRB correlations  \citep[see, e.g.,][]{2021JCAP...09..042K}.
\item[3)] Comparing the values of $\Delta\ln B$, we notice that the case $\Omega_k=0$ is statistically favored over the curved one, which is always consistent with $\Omega_k\approx0$. Moreover, using $\Omega_k=0$ provides better accuracy on the DE parameters.
\item[4)] In all the analyses where the spatial curvature is accounted for, $H_0$ decreases by about $\sim0.5\%$ and, thus, it can be considered weakly independent from $\Omega_k$, in line with the conclusions of \citet{2022JHEAp..33...10Z}. 
However, it is particularly evident comparing the outcomes of B1 analysis that accounting for $\Omega_k$ accommodates an overall shift of all the values of $\omega_i$ towards the cosmological constant value $\omega=-1$ at redshifts beyond $z\sim0.5$. 
At this level, we cannot disentangle between spatial curvature or dynamical DE as the reason for this $0.5\%$ reduction in $H_0$ (for similar conclusions, see Muccino \& Luongo in publication).
\item[5)] At $z\leq0.55$, all the analyses confirm a departure $\gtrsim2\sigma$ with respect to the value $\omega=-1$ of the $\Lambda$CDM model, in contrast with previous findings \citep{2021ApJ...908..181M}.
\item[6)] Conversely, at $z>0.55$, in all the analyses the DE parameters are compatible within $\sim1\sigma$ with the $\Lambda$CDM model, in contrast with previous findings \citep{2021ApJ...908..181M}. This is more evident when examining B2 results, also favored from a statistical standpoint by comparing $\Delta \ln B$ values. 
\end{itemize}

Focusing on the outcomes of the DE parametric reconstruction, we see that (a) the improvements  introduced by the new X-ray afterglow fitting procedure (see Sec.~\ref{sec:2}) and (b) the increased sample from $174$ to $244$ sources, especially at large redshifts (see Table~\ref{tab:no}), are particularly helpful in deriving constraints at $z>0.55$.
Nevertheless, at $z\leq0.55$, the $\Lambda$CDM model seems to fail in fitting GRB data, hinting at a possible dynamical behavior of DE at low redshifts.
This seems in line with the results of \citet{2022MNRAS.516.2575J} where the same parametric DE reconstruction method -- but involving GRBs, CMB, \emph{Pantheon} SNe Ia, and cosmic chronometers -- an oscillating behavior around $\omega=-1$ was reported at $z\lesssim0.5$.
However, back to our analysis involving GRB data alone, the results from the redshift bin at $z\leq0.55$ certify that, albeit performing better, the C244 sample still suffers from the issues listed below.
\begin{itemize}
\item[-] The shortage of low-$z$ GRBs severely influences our results at $z \lesssim 0.5$, where the DE effect starts to dominate but only $\sim 10\%$ of the Combo-GRBs are located, indicating a lack of statistics.
This shortage of nearby events is a direct consequence of the long GRB formation rate, observationally peaking at $z=2$ and declining at lower redshift, in broad agreement with the evolution of the cosmic star formation rate \citep[see, e.g.,][]{Pescalli:2015yva}.
\item[-] The lowest-redshift Combo-GRBs are located in the left-bottom corner of the correlation plane, where the best-fit correlation line seems to overestimate $L_0$ for these bursts (see Fig.~\ref{fig:no}) and, thus, the corresponding $\mu$. As a consequence, the analyses provide $\omega<-1$. 
This issue might be solved using the LLH proposed by \citet{Reichart2001}, rather than the one proposed by \citet{Dago2005}, hereby utilized in our analyses.
\end{itemize}

In view of the above, considering the analysis at $z\leq0.55$ to be inconclusive, in general, the parametrization indicates no dynamical DE and favors the cosmological constant.

\section{Discussion}\label{sec:5}

In this work, we considered the updated Combo sample, the C244, based on $244$ long GRBs. In particular, we:
\begin{itemize}
\item updated and verified all the redshifts and the prompt peak energies $E_{\rm p}$ of the sample,
\item performed new fits of the X-ray afterglows that refined the afterglow observables, and
\item improved the calibration jointly employing $11$ GRB sub-samples within the range $0.54\lesssim z \lesssim 3.59$ -- to determine slope and extra-scatter parameters -- together with three catalogs of SNe Ia at redshifts $z\leq0.075$ -- to constrain the intercept -- and the B\'ezier interpolation of the Hubble rate, since cosmographic series fail to be constraining up to $z \approx 3.59$. 
\end{itemize}
In view of the above, we investigated the possibility to extract cosmological bounds using only GRBs.
In particular, we focused on the recent revival of the dynamical nature of DE.
This hot topic was recently triggered by the results of DESI \citep{2025arXiv250314738D}, which exhibited a tension with the $\Lambda$CDM model \citep{Planck2018} and the preference for a CPL dynamical DE model with $w_0>−1$ and $w_{\rm a}<0$.

We tackled the problem in a model-independent way constraining the evolution of DE EoS parameter through a piecewise redshift-binned formulation.
We considered two analyses: B1, based on $5$ redshift bins, and B2, based on an averaged parametrization over only two bins. Further, we also investigated the effects of the spatial curvature.
The main results of our DE EoS reconstruction analyses are summarized below.
\begin{itemize}
\item The statistical analysis tends to favor the results obtained by considering spatial flatness. This is also confirmed by the best-fit values of $\Omega_m$, more consistent with the value purported by the $\Lambda$CDM model, and by the better accuracy on cosmological parameters.
\item Although statistically excluded, the B1 analysis with $\Omega_k\neq0$ leads to a decrease of $H_0$ of about $\sim0.5\%$ (with respect to the flat scenario) and induces, at $z\gtrsim0.5$, an overall shift of $\omega_i$ towards the standard value $\omega=-1$.
This does not exclude a possible interplay between spatial curvature and dynamical DE.
\item At $z\leq0.55$, all the analyses produced values of the DE parameter $\omega<-1$, with a $\gtrsim2\sigma$ departure with respect to $\omega=-1$. However, this phantom behavior of DE is likely the byproduct of the shortage of GRBs at $z \lesssim 0.55$. For these reasons, the analysis at $z\leq0.55$ is considered inconclusive.
\item At $z>0.55$, all the analyses -- in particular B2 -- provided DE parameters compatible within $\sim1\sigma$ with the standard $\Lambda$CDM model. Contrary to the previous case, the results on the DE parametrization at high redshifts are statistically more robust and indicate no dynamical DE behavior, favoring the cosmological constant case predicted by the $\Lambda$CDM model.
\end{itemize}

\section{Future prospects}\label{sec:6}

From the above considerations, it is clear that opportunely standardized and tighter GRB correlations are crucial for investigating the evolution of the universe at high redshifts, inaccessible to most of the astrophysical probes. 
However, competitive constraints are in general jeopardized by the shortage of low-redshift events, including the low-luminosity bursts like GRB~060218, GRB~120422, and GRB~171205. A possible solution to populate this redshift interval would consist of verifying whether the Combo correlation is fulfilled by fast X-ray transient (FXT) events, recently and jointly observed by SVOM (Space Variable Objects Monitor, \citealt{SVOM}) and EP (Einstein Probe, \citealt{EinsteinProbeTeam:2015bcj}) missions. 
FXTs are intense, short-lived bursts of X-ray photons lasting from a few minutes to several hours that might be either high-redshift GRBs or intrinsically weak events, e.g., off-axis events or shock breakout dominated GRB--SNe connections.
In these respects, future missions like THESEUS \citep{2021ExA....52..183A} will strengthen the importance of GRBs, enabling robust detections of bursts and more detailed study of the early universe up the reionization era \citep{2024arXiv241220424L}.
The synergy with surveys such as Euclid \citep{2025A&A...697A...1E} will provide additional data and further constraints on the still open puzzle of DE EoS.

\section{Conclusions}\label{sec:7}

Overall, the results presented in this work provide a coherent picture that can be summarized in the following key conclusions.
\begin{itemize}
\item[1)] GRBs show no significant evidence for an evolving EoS $\omega(z)$ of DE with the redshift. The apparent ``phantom'' behavior at $z\lesssim0.55$ mainly reflects the shortage of nearby GRBs and, thus, the lack of statistics rather than representing the evidence for dynamical DE.
\item[2)] The calibration of the Combo correlation yields $H_0\sim70$~km/s/Mpc, consistent with the value determined by  \citet{2021A&A...647A..72K}. This value alleviates the existing $H_0$ tension and it is broadly consistent with current measurements based on the CMB (at $\approx1.3$-sigma, \citealt{Planck2018}) and Cepheid/SNe~Ia (at $\lesssim1$-sigma,  \citealt{2022ApJ...934L...7R}), although the uncertainties do not allow a competitive measurement. 
\item[3)] GRBs, as distance indicators, are broadly consistent with the current cosmic distance ladder.
\end{itemize}

\begin{acknowledgements}
This work made use of data supplied by the UK \emph{Swift} Science Data Center at the University of Leicester.
\end{acknowledgements}

\bibliographystyle{aa}
\bibliography{biblio}

\begin{appendix}

\onecolumn

\section{The C244 catalog}\label{appendix1}

Table~\ref{tab:no} lists the handful observables for the C244 catalog and the deduced standardized distance moduli.
\scriptsize{
\LTcapwidth=\linewidth
\setlength{\tabcolsep}{1.5em}
\renewcommand{\arraystretch}{1.}
\begin{longtable}{llcrrrrr}
\caption{The GRBs of the C244 catalog with calibration sub-samples, correlation observables and standardized distance moduli.} 
\label{tab:no}\\
\hline\hline
GRB  &  z  &  Sub-sample  &  $\log E_{\rm p}$  &  $\log F_0$  &  $\log T$  &  $\mu_{\rm GRB}(\Omega_k=0)$  &  $\mu_{\rm GRB}(\Omega_k\neq0)$ \\
\hline
\endfirsthead
\caption{Continued.}\\
\hline\hline
GRB  &  z  &  Sub-sample  &  $\log E_{\rm p}$  &  $\log F_0$  &  $\log T$  &  $\mu_{\rm GRB}(\Omega_k=0)$  &  $\mu_{\rm GRB}(\Omega_k\neq0)$ \\
\hline
\endhead
\hline
\endfoot
050215B	&	2.62	&	9	& $	1.80	\pm	0.26	$ & $	-11.76	\pm	0.18	$ & $	4.09	\pm	0.41	$ & $	47.42	\pm	1.72	$ & $	47.21	\pm	1.84	$	\\
050315A	&	1.949	&	 	& $	2.07	\pm	0.11	$ & $	-10.82	\pm	0.02	$ & $	4.55	\pm	0.08	$ & $	44.54	\pm	1.23	$ & $	44.30	\pm	1.42	$	\\
050318A	&	1.44	&	5	& $	2.06	\pm	0.10	$ & $	-10.03	\pm	0.09	$ & $	3.30	\pm	0.14	$ & $	45.63	\pm	1.27	$ & $	45.39	\pm	1.46	$	\\
050401A	&	2.9	&	10	& $	2.67	\pm	0.10	$ & $	-9.32	\pm	0.01	$ & $	2.86	\pm	0.04	$ & $	46.38	\pm	1.28	$ & $	46.08	\pm	1.53	$	\\
050416A	&	0.6535	&	2	& $	1.34	\pm	0.09	$ & $	-10.18	\pm	0.05	$ & $	4.34	\pm	0.08	$ & $	41.78	\pm	1.18	$ & $	41.63	\pm	1.30	$	\\
050505A	&	4.27	&	 	& $	2.82	\pm	0.17	$ & $	-10.31	\pm	0.03	$ & $	3.43	\pm	0.05	$ & $	47.78	\pm	1.34	$ & $	47.47	\pm	1.59	$	\\
050525A	&	0.606	&	2	& $	2.11	\pm	0.02	$ & $	-9.68	\pm	0.22	$ & $	3.38	\pm	0.20	$ & $	44.70	\pm	1.39	$ & $	44.46	\pm	1.57	$	\\
050603A	&	2.821	&	 	& $	3.12	\pm	0.03	$ & $	-7.96	\pm	0.27	$ & $	2.00	\pm	0.18	$ & $	46.16	\pm	1.56	$ & $	45.82	\pm	1.81	$	\\
050814A	&	5.77	&	 	& $	2.56	\pm	0.06	$ & $	-11.29	\pm	0.05	$ & $	3.69	\pm	0.12	$ & $	48.97	\pm	1.29	$ & $	48.69	\pm	1.52	$	\\
050820A	&	2.612	&	9	& $	3.12	\pm	0.09	$ & $	-9.73	\pm	0.03	$ & $	3.61	\pm	0.04	$ & $	46.54	\pm	1.35	$ & $	46.20	\pm	1.63	$	\\
050904A	&	6.29	&	 	& $	3.50	\pm	0.16	$ & $	-10.47	\pm	0.16	$ & $	2.78	\pm	0.20	$ & $	51.36	\pm	1.58	$ & $	50.97	\pm	1.85	$	\\
050922C	&	2.198	&	8	& $	2.62	\pm	0.12	$ & $	-8.66	\pm	0.07	$ & $	2.07	\pm	0.09	$ & $	46.58	\pm	1.30	$ & $	46.29	\pm	1.54	$	\\
051008A	&	2.77	&	 	& $	3.33	\pm	0.18	$ & $	-10.22	\pm	0.04	$ & $	3.33	\pm	0.06	$ & $	48.97	\pm	1.44	$ & $	48.60	\pm	1.72	$	\\
051016B	&	0.9364	&	 	& $	1.73	\pm	0.23	$ & $	-11.07	\pm	0.04	$ & $	4.23	\pm	0.10	$ & $	45.17	\pm	1.29	$ & $	44.97	\pm	1.44	$	\\
051022A	&	0.809	&	 	& $	2.88	\pm	0.15	$ & $	-9.85	\pm	0.09	$ & $	3.96	\pm	0.10	$ & $	45.44	\pm	1.37	$ & $	45.12	\pm	1.62	$	\\
051109A	&	2.346	&	 	& $	2.73	\pm	0.17	$ & $	-9.82	\pm	0.08	$ & $	3.45	\pm	0.10	$ & $	46.30	\pm	1.36	$ & $	45.99	\pm	1.60	$	\\
060111A	&	2.32	&	8	& $	2.38	\pm	0.07	$ & $	-11.30	\pm	0.10	$ & $	4.13	\pm	0.18	$ & $	47.49	\pm	1.33	$ & $	47.23	\pm	1.54	$	\\
060115A	&	3.53	&	11	& $	2.45	\pm	0.05	$ & $	-10.91	\pm	0.07	$ & $	3.74	\pm	0.20	$ & $	47.66	\pm	1.34	$ & $	47.38	\pm	1.56	$	\\
060124A	&	2.296	&	8	& $	2.89	\pm	0.17	$ & $	-9.78	\pm	0.02	$ & $	3.76	\pm	0.03	$ & $	45.80	\pm	1.35	$ & $	45.48	\pm	1.60	$	\\
060204B	&	2.3393	&	8	& $	2.51	\pm	0.20	$ & $	-10.28	\pm	0.07	$ & $	3.40	\pm	0.11	$ & $	47.04	\pm	1.35	$ & $	46.76	\pm	1.57	$	\\
060502A	&	1.51	&	5	& $	2.51	\pm	0.14	$ & $	-10.06	\pm	0.07	$ & $	4.02	\pm	0.10	$ & $	44.95	\pm	1.31	$ & $	44.67	\pm	1.53	$	\\
060206A	&	4.048	&	 	& $	2.60	\pm	0.05	$ & $	-9.96	\pm	0.06	$ & $	3.46	\pm	0.07	$ & $	46.31	\pm	1.27	$ & $	46.02	\pm	1.51	$	\\
060210A	&	3.91	&	 	& $	2.76	\pm	0.15	$ & $	-9.96	\pm	0.04	$ & $	3.50	\pm	0.06	$ & $	46.56	\pm	1.32	$ & $	46.25	\pm	1.57	$	\\
060218A	&	0.0331	&	 	& $	0.69	\pm	0.03	$ & $	-10.64	\pm	0.21	$ & $	5.48	\pm	0.41	$ & $	38.57	\pm	1.62	$ & $	38.48	\pm	1.66	$	\\
060306A	&	1.559	&	 	& $	2.25	\pm	0.20	$ & $	-10.28	\pm	0.04	$ & $	3.59	\pm	0.07	$ & $	45.99	\pm	1.30	$ & $	45.73	\pm	1.50	$	\\
060418A	&	1.489	&	5	& $	2.76	\pm	0.11	$ & $	-8.46	\pm	0.17	$ & $	2.12	\pm	0.18	$ & $	46.27	\pm	1.43	$ & $	45.97	\pm	1.66	$	\\
060512A	&	2.1	&	7	& $	1.85	\pm	0.37	$ & $	-10.62	\pm	0.63	$ & $	3.32	\pm	0.90	$ & $	46.60	\pm	3.10	$ & $	46.39	\pm	3.17	$	\\
060522A	&	5.11	&	 	& $	2.63	\pm	0.08	$ & $	-10.61	\pm	0.37	$ & $	3.10	\pm	1.14	$ & $	48.92	\pm	3.25	$ & $	48.62	\pm	3.35	$	\\
060526A	&	3.221	&	 	& $	2.02	\pm	0.09	$ & $	-11.00	\pm	0.06	$ & $	3.71	\pm	0.21	$ & $	46.94	\pm	1.31	$ & $	46.71	\pm	1.49	$	\\
060604A	&	2.1357	&	 	& $	2.10	\pm	0.05	$ & $	-11.04	\pm	0.05	$ & $	4.03	\pm	0.10	$ & $	46.42	\pm	1.22	$ & $	46.18	\pm	1.42	$	\\
060605A	&	3.78	&	 	& $	2.69	\pm	0.26	$ & $	-10.15	\pm	0.04	$ & $	3.01	\pm	0.07	$ & $	48.14	\pm	1.40	$ & $	47.84	\pm	1.62	$	\\
060607A	&	3.0749	&	10	& $	2.68	\pm	0.11	$ & $	-9.71	\pm	0.07	$ & $	2.86	\pm	0.16	$ & $	47.38	\pm	1.35	$ & $	47.08	\pm	1.58	$	\\
060707A	&	3.425	&	 	& $	2.45	\pm	0.04	$ & $	-10.45	\pm	0.07	$ & $	4.04	\pm	0.13	$ & $	45.73	\pm	1.28	$ & $	45.46	\pm	1.51	$	\\
060708A	&	1.92	&	 	& $	2.47	\pm	0.26	$ & $	-9.99	\pm	0.07	$ & $	3.09	\pm	0.08	$ & $	47.02	\pm	1.38	$ & $	46.74	\pm	1.59	$	\\
060714A	&	2.711	&	9	& $	2.37	\pm	0.23	$ & $	-10.38	\pm	0.04	$ & $	3.37	\pm	0.08	$ & $	47.05	\pm	1.34	$ & $	46.79	\pm	1.54	$	\\
060729A	&	0.54	&	1	& $	1.89	\pm	0.23	$ & $	-10.48	\pm	0.01	$ & $	4.91	\pm	0.02	$ & $	42.34	\pm	1.28	$ & $	42.13	\pm	1.45	$	\\
060814A	&	1.9229	&	 	& $	2.88	\pm	0.15	$ & $	-10.33	\pm	0.05	$ & $	3.90	\pm	0.08	$ & $	46.78	\pm	1.35	$ & $	46.46	\pm	1.60	$	\\
060906A	&	3.686	&	11	& $	2.32	\pm	0.09	$ & $	-10.95	\pm	0.05	$ & $	3.21	\pm	0.12	$ & $	48.76	\pm	1.27	$ & $	48.50	\pm	1.48	$	\\
060908A	&	1.8836	&	 	& $	2.64	\pm	0.09	$ & $	-8.86	\pm	0.09	$ & $	2.38	\pm	0.14	$ & $	46.35	\pm	1.33	$ & $	46.06	\pm	1.56	$	\\
060927A	&	5.467	&	 	& $	2.68	\pm	0.04	$ & $	-10.14	\pm	0.06	$ & $	2.25	\pm	0.13	$ & $	49.97	\pm	1.31	$ & $	49.67	\pm	1.55	$	\\
061007A	&	1.261	&	4	& $	2.95	\pm	0.06	$ & $	-6.43	\pm	0.18	$ & $	1.14	\pm	0.12	$ & $	44.10	\pm	1.41	$ & $	43.78	\pm	1.66	$	\\
061021A	&	0.3463	&	 	& $	2.85	\pm	0.20	$ & $	-9.84	\pm	0.05	$ & $	4.15	\pm	0.06	$ & $	44.87	\pm	1.37	$ & $	44.55	\pm	1.62	$	\\
061121A	&	1.314	&	 	& $	3.11	\pm	0.05	$ & $	-9.42	\pm	0.01	$ & $	3.47	\pm	0.02	$ & $	46.13	\pm	1.33	$ & $	45.78	\pm	1.61	$	\\
061126A	&	1.1588	&	3	& $	3.13	\pm	0.14	$ & $	-8.29	\pm	0.15	$ & $	2.50	\pm	0.13	$ & $	45.75	\pm	1.45	$ & $	45.41	\pm	1.71	$	\\
061222A	&	2.088	&	7	& $	2.94	\pm	0.08	$ & $	-9.46	\pm	0.02	$ & $	3.49	\pm	0.03	$ & $	45.80	\pm	1.31	$ & $	45.47	\pm	1.58	$	\\
070125A	&	1.547	&	 	& $	2.97	\pm	0.07	$ & $	-11.05	\pm	0.08	$ & $	4.42	\pm	0.15	$ & $	47.50	\pm	1.38	$ & $	47.17	\pm	1.64	$	\\
070129A	&	2.3384	&	8	& $	2.16	\pm	0.12	$ & $	-10.99	\pm	0.04	$ & $	4.28	\pm	0.09	$ & $	45.85	\pm	1.25	$ & $	45.61	\pm	1.45	$	\\
070208A	&	1.165	&	 	& $	2.05	\pm	0.01	$ & $	-11.20	\pm	0.12	$ & $	3.83	\pm	0.31	$ & $	47.23	\pm	1.44	$ & $	46.99	\pm	1.61	$	\\
070306A	&	1.4959	&	5	& $	2.36	\pm	0.08	$ & $	-10.40	\pm	0.02	$ & $	4.16	\pm	0.06	$ & $	45.10	\pm	1.24	$ & $	44.84	\pm	1.46	$	\\
070328A	&	2.0627	&	7	& $	3.18	\pm	0.08	$ & $	-8.58	\pm	0.01	$ & $	2.55	\pm	0.03	$ & $	46.48	\pm	1.35	$ & $	46.13	\pm	1.63	$	\\
070521A	&	2.0865	&	7	& $	2.84	\pm	0.05	$ & $	-9.44	\pm	0.04	$ & $	3.02	\pm	0.06	$ & $	46.67	\pm	1.29	$ & $	46.35	\pm	1.56	$	\\
071003A	&	1.60435	&	 	& $	3.32	\pm	0.06	$ & $	-10.44	\pm	0.18	$ & $	3.83	\pm	0.20	$ & $	48.22	\pm	1.53	$ & $	47.85	\pm	1.80	$	\\
071020A	&	2.145	&	 	& $	3.01	\pm	0.07	$ & $	-8.51	\pm	0.07	$ & $	1.87	\pm	0.09	$ & $	47.60	\pm	1.35	$ & $	47.26	\pm	1.61	$	\\
071031A	&	2.692	&	9	& $	1.95	\pm	0.13	$ & $	-11.86	\pm	0.13	$ & $	4.08	\pm	0.38	$ & $	48.03	\pm	1.57	$ & $	47.80	\pm	1.72	$	\\
071117A	&	1.331	&	 	& $	2.05	\pm	0.24	$ & $	-8.85	\pm	0.02	$ & $	2.71	\pm	0.14	$ & $	44.14	\pm	1.35	$ & $	43.91	\pm	1.52	$	\\
080207A	&	2.0858	&	7	& $	2.52	\pm	0.35	$ & $	-8.57	\pm	0.02	$ & $	2.51	\pm	0.07	$ & $	45.04	\pm	1.48	$ & $	44.75	\pm	1.67	$	\\
080319B	&	0.937	&	 	& $	3.10	\pm	0.02	$ & $	-7.26	\pm	0.02	$ & $	2.22	\pm	0.02	$ & $	43.80	\pm	1.33	$ & $	43.46	\pm	1.61	$	\\
080319C	&	1.95	&	 	& $	2.96	\pm	0.13	$ & $	-8.63	\pm	0.04	$ & $	2.83	\pm	0.07	$ & $	45.39	\pm	1.35	$ & $	45.06	\pm	1.61	$	\\
080411A	&	1.03	&	 	& $	2.72	\pm	0.06	$ & $	-9.10	\pm	0.05	$ & $	3.65	\pm	0.05	$ & $	43.96	\pm	1.28	$ & $	43.65	\pm	1.53	$	\\
080413A	&	2.433	&	9	& $	2.77	\pm	0.14	$ & $	-9.33	\pm	0.14	$ & $	2.26	\pm	0.26	$ & $	48.12	\pm	1.51	$ & $	47.81	\pm	1.73	$	\\
080413B	&	1.1	&	3	& $	2.21	\pm	0.13	$ & $	-9.21	\pm	0.03	$ & $	3.17	\pm	0.06	$ & $	44.28	\pm	1.24	$ & $	44.03	\pm	1.45	$	\\
080605A	&	1.6398	&	 	& $	2.81	\pm	0.04	$ & $	-8.42	\pm	0.02	$ & $	2.32	\pm	0.03	$ & $	45.81	\pm	1.28	$ & $	45.49	\pm	1.54	$	\\
080603B	&	2.69	&	9	& $	2.58	\pm	0.12	$ & $	-9.63	\pm	0.12	$ & $	3.19	\pm	0.53	$ & $	46.11	\pm	1.85	$ & $	45.83	\pm	2.02	$	\\
080607A	&	3.036	&	10	& $	3.23	\pm	0.06	$ & $	-8.93	\pm	0.04	$ & $	2.36	\pm	0.06	$ & $	47.93	\pm	1.36	$ & $	47.57	\pm	1.65	$	\\
080721A	&	2.591	&	9	& $	3.24	\pm	0.06	$ & $	-8.16	\pm	0.01	$ & $	2.31	\pm	0.01	$ & $	46.14	\pm	1.36	$ & $	45.79	\pm	1.64	$	\\
080804A	&	2.2045	&	8	& $	2.85	\pm	0.05	$ & $	-9.19	\pm	0.09	$ & $	2.19	\pm	0.11	$ & $	48.12	\pm	1.33	$ & $	47.81	\pm	1.59	$	\\
080905B	&	2.374	&	 	& $	2.79	\pm	0.12	$ & $	-9.85	\pm	0.03	$ & $	3.38	\pm	0.06	$ & $	46.68	\pm	1.31	$ & $	46.37	\pm	1.57	$	\\
080810A	&	3.35	&	 	& $	3.17	\pm	0.05	$ & $	-9.66	\pm	0.08	$ & $	2.88	\pm	0.09	$ & $	48.34	\pm	1.38	$ & $	47.99	\pm	1.66	$	\\
080913A	&	6.695	&	 	& $	2.85	\pm	0.23	$ & $	-10.49	\pm	0.19	$ & $	2.96	\pm	0.31	$ & $	49.47	\pm	1.66	$ & $	49.16	\pm	1.86	$	\\
080916A	&	0.689	&	 	& $	2.25	\pm	0.09	$ & $	-10.06	\pm	0.07	$ & $	4.04	\pm	0.11	$ & $	44.31	\pm	1.26	$ & $	44.06	\pm	1.47	$	\\
080928A	&	1.692	&	6	& $	1.98	\pm	0.11	$ & $	-9.97	\pm	0.18	$ & $	3.12	\pm	0.21	$ & $	45.77	\pm	1.38	$ & $	45.54	\pm	1.55	$	\\
081007A	&	0.5295	&	1	& $	1.79	\pm	0.11	$ & $	-9.92	\pm	0.05	$ & $	4.02	\pm	0.10	$ & $	42.96	\pm	1.22	$ & $	42.75	\pm	1.38	$	\\
081008A	&	1.9685	&	 	& $	2.42	\pm	0.09	$ & $	-10.13	\pm	0.04	$ & $	3.19	\pm	0.08	$ & $	46.99	\pm	1.25	$ & $	46.72	\pm	1.48	$	\\
081028A	&	3.038	&	10	& $	2.37	\pm	0.18	$ & $	-11.33	\pm	0.05	$ & $	4.05	\pm	0.15	$ & $	47.75	\pm	1.34	$ & $	47.49	\pm	1.55	$	\\
081109A	&	0.9787	&	 	& $	1.75	\pm	0.18	$ & $	-9.05	\pm	0.04	$ & $	2.84	\pm	0.06	$ & $	43.65	\pm	1.24	$ & $	43.45	\pm	1.40	$	\\
081118A	&	2.58	&	9	& $	2.17	\pm	0.04	$ & $	-11.46	\pm	0.12	$ & $	4.89	\pm	0.27	$ & $	45.49	\pm	1.40	$ & $	45.25	\pm	1.59	$	\\
081121A	&	2.512	&	9	& $	2.94	\pm	0.06	$ & $	-9.20	\pm	0.08	$ & $	3.00	\pm	0.08	$ & $	46.34	\pm	1.33	$ & $	46.01	\pm	1.60	$	\\
081203A	&	2.05	&	7	& $	3.19	\pm	0.23	$ & $	-9.08	\pm	0.02	$ & $	2.56	\pm	0.05	$ & $	47.71	\pm	1.45	$ & $	47.36	\pm	1.71	$	\\
081221A	&	2.26	&	8	& $	2.45	\pm	0.01	$ & $	-8.47	\pm	0.11	$ & $	2.40	\pm	0.14	$ & $	44.89	\pm	1.30	$ & $	44.61	\pm	1.53	$	\\
081222A	&	2.77	&	 	& $	2.70	\pm	0.03	$ & $	-8.67	\pm	0.03	$ & $	2.32	\pm	0.04	$ & $	46.18	\pm	1.27	$ & $	45.88	\pm	1.52	$	\\
090102A	&	1.547	&	 	& $	3.06	\pm	0.06	$ & $	-8.26	\pm	0.08	$ & $	2.20	\pm	0.08	$ & $	46.26	\pm	1.36	$ & $	45.92	\pm	1.63	$	\\
090113A	&	1.7493	&	6	& $	2.59	\pm	0.10	$ & $	-9.30	\pm	0.08	$ & $	2.63	\pm	0.12	$ & $	46.73	\pm	1.31	$ & $	46.44	\pm	1.54	$	\\
090205A	&	4.6497	&	 	& $	2.33	\pm	0.16	$ & $	-10.78	\pm	0.04	$ & $	3.11	\pm	0.12	$ & $	48.61	\pm	1.30	$ & $	48.35	\pm	1.51	$	\\
090418A	&	1.608	&	 	& $	3.20	\pm	0.11	$ & $	-9.35	\pm	0.02	$ & $	3.00	\pm	0.04	$ & $	47.31	\pm	1.37	$ & $	46.96	\pm	1.65	$	\\
090423A	&	8.26	&	 	& $	2.69	\pm	0.19	$ & $	-11.25	\pm	0.06	$ & $	3.14	\pm	0.13	$ & $	50.54	\pm	1.38	$ & $	50.24	\pm	1.61	$	\\
090429B	&	9.4	&	 	& $	2.65	\pm	0.06	$ & $	-10.79	\pm	0.05	$ & $	2.72	\pm	0.13	$ & $	50.35	\pm	1.31	$ & $	50.05	\pm	1.55	$	\\
090424A	&	0.544	&	1	& $	2.37	\pm	0.01	$ & $	-8.26	\pm	0.02	$ & $	3.01	\pm	0.03	$ & $	42.67	\pm	1.22	$ & $	42.40	\pm	1.45	$	\\
090516A	&	4.109	&	 	& $	2.99	\pm	0.18	$ & $	-10.20	\pm	0.12	$ & $	3.20	\pm	0.17	$ & $	48.45	\pm	1.47	$ & $	48.12	\pm	1.71	$	\\
090530A	&	1.266	&	4	& $	2.32	\pm	0.17	$ & $	-10.52	\pm	0.05	$ & $	4.04	\pm	0.10	$ & $	45.62	\pm	1.30	$ & $	45.36	\pm	1.51	$	\\
090618A	&	0.54	&	1	& $	2.36	\pm	0.01	$ & $	-8.87	\pm	0.01	$ & $	3.42	\pm	0.02	$ & $	43.15	\pm	1.21	$ & $	42.88	\pm	1.44	$	\\
090715B	&	3.	&	10	& $	2.73	\pm	0.14	$ & $	-9.72	\pm	0.17	$ & $	2.83	\pm	0.16	$ & $	47.58	\pm	1.43	$ & $	47.28	\pm	1.66	$	\\
090812A	&	2.452	&	9	& $	3.30	\pm	0.16	$ & $	-9.50	\pm	0.09	$ & $	2.97	\pm	0.19	$ & $	47.99	\pm	1.50	$ & $	47.63	\pm	1.77	$	\\
090927A	&	1.37	&	 	& $	2.67	\pm	0.16	$ & $	-11.27	\pm	0.08	$ & $	4.09	\pm	0.19	$ & $	48.17	\pm	1.40	$ & $	47.87	\pm	1.63	$	\\
091018A	&	0.971	&	 	& $	1.74	\pm	0.17	$ & $	-8.97	\pm	0.03	$ & $	2.77	\pm	0.05	$ & $	43.59	\pm	1.22	$ & $	43.39	\pm	1.38	$	\\
091020A	&	1.71	&	6	& $	2.45	\pm	0.36	$ & $	-9.20	\pm	0.02	$ & $	2.82	\pm	0.05	$ & $	45.66	\pm	1.48	$ & $	45.39	\pm	1.66	$	\\
091029A	&	2.752	&	 	& $	2.36	\pm	0.13	$ & $	-10.78	\pm	0.03	$ & $	3.94	\pm	0.06	$ & $	46.61	\pm	1.26	$ & $	46.34	\pm	1.48	$	\\
091127A	&	0.49	&	 	& $	1.78	\pm	0.01	$ & $	-8.99	\pm	0.03	$ & $	3.65	\pm	0.03	$ & $	41.54	\pm	1.16	$ & $	41.34	\pm	1.34	$	\\
091208B	&	1.063	&	 	& $	2.31	\pm	0.04	$ & $	-9.72	\pm	0.04	$ & $	3.30	\pm	0.07	$ & $	45.43	\pm	1.23	$ & $	45.18	\pm	1.45	$	\\
100302A	&	4.813	&	 	& $	2.72	\pm	0.18	$ & $	-11.75	\pm	0.08	$ & $	5.11	\pm	0.21	$ & $	46.94	\pm	1.44	$ & $	46.64	\pm	1.67	$	\\
100413A	&	3.9	&	 	& $	3.25	\pm	0.09	$ & $	-9.10	\pm	0.04	$ & $	2.58	\pm	0.06	$ & $	47.85	\pm	1.38	$ & $	47.49	\pm	1.66	$	\\
100418A	&	0.6235	&	2	& $	1.67	\pm	0.03	$ & $	-11.82	\pm	0.05	$ & $	5.19	\pm	0.18	$ & $	44.49	\pm	1.25	$ & $	44.30	\pm	1.40	$	\\
100615A	&	1.398	&	 	& $	2.11	\pm	0.06	$ & $	-9.37	\pm	0.04	$ & $	3.66	\pm	0.08	$ & $	43.23	\pm	1.21	$ & $	42.99	\pm	1.41	$	\\
100621A	&	0.542	&	1	& $	2.17	\pm	0.07	$ & $	-9.28	\pm	0.04	$ & $	4.45	\pm	0.07	$ & $	41.16	\pm	1.22	$ & $	40.91	\pm	1.43	$	\\
100704A	&	3.6	&	11	& $	2.91	\pm	0.07	$ & $	-10.15	\pm	0.05	$ & $	3.70	\pm	0.08	$ & $	46.89	\pm	1.32	$ & $	46.57	\pm	1.59	$	\\
100728A	&	1.567	&	 	& $	2.90	\pm	0.01	$ & $	-9.06	\pm	0.04	$ & $	3.20	\pm	0.05	$ & $	45.43	\pm	1.30	$ & $	45.10	\pm	1.57	$	\\
100728B	&	2.106	&	7	& $	2.51	\pm	0.06	$ & $	-9.63	\pm	0.05	$ & $	2.48	\pm	0.10	$ & $	47.75	\pm	1.27	$ & $	47.47	\pm	1.50	$	\\
100814A	&	1.44	&	5	& $	2.49	\pm	0.07	$ & $	-10.56	\pm	0.04	$ & $	4.36	\pm	0.07	$ & $	45.33	\pm	1.25	$ & $	45.05	\pm	1.49	$	\\
100906A	&	1.727	&	6	& $	2.59	\pm	0.32	$ & $	-9.60	\pm	0.02	$ & $	3.23	\pm	0.04	$ & $	45.97	\pm	1.45	$ & $	45.68	\pm	1.65	$	\\
101213A	&	0.414	&	 	& $	2.69	\pm	0.04	$ & $	-9.62	\pm	0.06	$ & $	4.40	\pm	0.11	$ & $	43.31	\pm	1.30	$ & $	43.01	\pm	1.55	$	\\
101219B	&	0.5519	&	1	& $	1.99	\pm	0.11	$ & $	-11.64	\pm	0.09	$ & $	6.06	\pm	0.20	$ & $	42.60	\pm	1.32	$ & $	42.38	\pm	1.50	$	\\
110106B	&	0.618	&	2	& $	2.33	\pm	0.05	$ & $	-10.00	\pm	0.05	$ & $	3.72	\pm	0.10	$ & $	45.16	\pm	1.24	$ & $	44.90	\pm	1.46	$	\\
110205A	&	2.22	&	8	& $	2.87	\pm	0.20	$ & $	-8.96	\pm	0.06	$ & $	2.46	\pm	0.07	$ & $	46.95	\pm	1.38	$ & $	46.63	\pm	1.63	$	\\
110213A	&	1.46	&	5	& $	2.35	\pm	0.14	$ & $	-9.13	\pm	0.02	$ & $	3.24	\pm	0.03	$ & $	44.21	\pm	1.25	$ & $	43.95	\pm	1.47	$	\\
110422A	&	1.77	&	6	& $	2.62	\pm	0.01	$ & $	-9.07	\pm	0.03	$ & $	3.02	\pm	0.05	$ & $	45.23	\pm	1.25	$ & $	44.93	\pm	1.50	$	\\
110503A	&	1.613	&	 	& $	2.76	\pm	0.04	$ & $	-8.41	\pm	0.05	$ & $	2.32	\pm	0.06	$ & $	45.66	\pm	1.29	$ & $	45.35	\pm	1.54	$	\\
110715A	&	0.82	&	 	& $	2.34	\pm	0.04	$ & $	-8.53	\pm	0.04	$ & $	2.89	\pm	0.06	$ & $	43.57	\pm	1.22	$ & $	43.30	\pm	1.45	$	\\
110731A	&	2.83	&	 	& $	3.07	\pm	0.02	$ & $	-8.04	\pm	0.09	$ & $	1.76	\pm	0.10	$ & $	46.83	\pm	1.36	$ & $	46.49	\pm	1.63	$	\\
110801A	&	1.858	&	 	& $	2.60	\pm	0.05	$ & $	-10.14	\pm	0.09	$ & $	3.36	\pm	0.14	$ & $	47.02	\pm	1.32	$ & $	46.73	\pm	1.55	$	\\
110818A	&	3.36	&	 	& $	3.05	\pm	0.10	$ & $	-9.98	\pm	0.26	$ & $	2.67	\pm	0.34	$ & $	49.36	\pm	1.71	$ & $	49.02	\pm	1.93	$	\\
111008A	&	4.9898	&	 	& $	2.95	\pm	0.12	$ & $	-10.38	\pm	0.03	$ & $	3.39	\pm	0.05	$ & $	48.33	\pm	1.33	$ & $	48.00	\pm	1.60	$	\\
111107A	&	2.893	&	10	& $	2.62	\pm	0.13	$ & $	-10.63	\pm	0.15	$ & $	3.13	\pm	0.28	$ & $	48.86	\pm	1.51	$ & $	48.57	\pm	1.72	$	\\
111123A	&	3.1516	&	 	& $	3.60	\pm	0.05	$ & $	-10.52	\pm	0.07	$ & $	3.59	\pm	0.17	$ & $	49.71	\pm	1.50	$ & $	49.31	\pm	1.79	$	\\
111209A	&	0.677	&	 	& $	2.72	\pm	0.07	$ & $	-11.01	\pm	0.16	$ & $	4.65	\pm	0.22	$ & $	46.25	\pm	1.44	$ & $	45.94	\pm	1.67	$	\\
111228A	&	0.714	&	 	& $	1.66	\pm	0.02	$ & $	-10.03	\pm	0.03	$ & $	4.07	\pm	0.05	$ & $	42.80	\pm	1.16	$ & $	42.61	\pm	1.32	$	\\
120118B	&	2.943	&	10	& $	2.23	\pm	0.03	$ & $	-10.67	\pm	0.04	$ & $	3.60	\pm	0.19	$ & $	46.90	\pm	1.29	$ & $	46.65	\pm	1.50	$	\\
120119A	&	1.728	&	6	& $	2.62	\pm	0.06	$ & $	-9.31	\pm	0.06	$ & $	2.79	\pm	0.09	$ & $	46.40	\pm	1.28	$ & $	46.11	\pm	1.52	$	\\
120326A	&	1.798	&	6	& $	2.11	\pm	0.03	$ & $	-10.96	\pm	0.05	$ & $	4.32	\pm	0.22	$ & $	45.56	\pm	1.31	$ & $	45.32	\pm	1.50	$	\\
120327A	&	2.813	&	 	& $	2.92	\pm	0.18	$ & $	-9.60	\pm	0.03	$ & $	2.83	\pm	0.08	$ & $	47.73	\pm	1.38	$ & $	47.40	\pm	1.63	$	\\
120422A	&	0.283	&	 	& $	1.52	\pm	0.57	$ & $	-12.31	\pm	0.07	$ & $	6.16	\pm	0.41	$ & $	42.97	\pm	2.02	$ & $	42.79	\pm	2.07	$	\\
120521C	&	6.	&	 	& $	2.86	\pm	0.21	$ & $	-11.04	\pm	0.08	$ & $	3.31	\pm	0.40	$ & $	49.99	\pm	1.71	$ & $	49.67	\pm	1.91	$	\\
120712A	&	4.1745	&	 	& $	2.81	\pm	0.09	$ & $	-9.51	\pm	0.03	$ & $	2.31	\pm	0.07	$ & $	48.56	\pm	1.30	$ & $	48.25	\pm	1.56	$	\\
120729A	&	0.8	&	 	& $	2.20	\pm	0.19	$ & $	-9.42	\pm	0.04	$ & $	2.92	\pm	0.07	$ & $	45.37	\pm	1.28	$ & $	45.13	\pm	1.48	$	\\
120811C	&	2.671	&	9	& $	2.20	\pm	0.06	$ & $	-9.79	\pm	0.06	$ & $	3.07	\pm	0.16	$ & $	45.95	\pm	1.27	$ & $	45.70	\pm	1.48	$	\\
120907A	&	0.97	&	 	& $	2.40	\pm	0.10	$ & $	-10.05	\pm	0.05	$ & $	3.51	\pm	0.08	$ & $	45.95	\pm	1.26	$ & $	45.68	\pm	1.49	$	\\
120909A	&	3.93	&	 	& $	3.22	\pm	0.03	$ & $	-9.51	\pm	0.08	$ & $	2.90	\pm	0.10	$ & $	48.03	\pm	1.39	$ & $	47.68	\pm	1.67	$	\\
120922A	&	3.1	&	 	& $	2.17	\pm	0.05	$ & $	-10.41	\pm	0.06	$ & $	3.60	\pm	0.10	$ & $	46.10	\pm	1.23	$ & $	45.85	\pm	1.44	$	\\
120923A	&	7.8	&	 	& $	2.59	\pm	0.11	$ & $	-11.06	\pm	0.17	$ & $	2.76	\pm	0.35	$ & $	50.80	\pm	1.59	$ & $	50.51	\pm	1.79	$	\\
121027A	&	1.773	&	6	& $	2.12	\pm	0.24	$ & $	-10.96	\pm	0.04	$ & $	4.80	\pm	0.10	$ & $	44.33	\pm	1.33	$ & $	44.09	\pm	1.51	$	\\
121128A	&	2.2	&	8	& $	2.39	\pm	0.02	$ & $	-8.88	\pm	0.03	$ & $	2.59	\pm	0.05	$ & $	45.30	\pm	1.22	$ & $	45.03	\pm	1.45	$	\\
121211A	&	1.023	&	 	& $	2.31	\pm	0.06	$ & $	-10.58	\pm	0.07	$ & $	4.01	\pm	0.15	$ & $	45.84	\pm	1.28	$ & $	45.58	\pm	1.50	$	\\
130408A	&	3.757	&	 	& $	3.00	\pm	0.06	$ & $	-9.95	\pm	0.05	$ & $	3.06	\pm	0.08	$ & $	48.22	\pm	1.33	$ & $	47.89	\pm	1.60	$	\\
130420A	&	1.297	&	4	& $	2.11	\pm	0.02	$ & $	-10.24	\pm	0.08	$ & $	4.58	\pm	0.14	$ & $	43.07	\pm	1.25	$ & $	42.84	\pm	1.45	$	\\
130427A	&	0.3399	&	 	& $	3.04	\pm	0.00	$ & $	-7.12	\pm	0.02	$ & $	2.82	\pm	0.03	$ & $	41.82	\pm	1.32	$ & $	41.49	\pm	1.59	$	\\
130505A	&	2.27	&	8	& $	3.31	\pm	0.02	$ & $	-9.34	\pm	0.03	$ & $	3.46	\pm	0.04	$ & $	46.40	\pm	1.37	$ & $	46.03	\pm	1.66	$	\\
130514A	&	3.6	&	11	& $	2.70	\pm	0.14	$ & $	-9.63	\pm	0.16	$ & $	2.85	\pm	0.17	$ & $	47.23	\pm	1.42	$ & $	46.93	\pm	1.65	$	\\
130606A	&	5.913	&	 	& $	3.31	\pm	0.11	$ & $	-9.43	\pm	0.15	$ & $	2.21	\pm	0.17	$ & $	49.76	\pm	1.50	$ & $	49.40	\pm	1.77	$	\\
130610A	&	2.092	&	7	& $	2.96	\pm	0.06	$ & $	-10.00	\pm	0.18	$ & $	2.77	\pm	0.22	$ & $	48.97	\pm	1.49	$ & $	48.64	\pm	1.73	$	\\
130612A	&	2.006	&	7	& $	2.27	\pm	0.07	$ & $	-10.13	\pm	0.24	$ & $	2.68	\pm	0.35	$ & $	47.91	\pm	1.61	$ & $	47.66	\pm	1.78	$	\\
130701A	&	1.155	&	3	& $	2.28	\pm	0.02	$ & $	-8.70	\pm	0.08	$ & $	2.34	\pm	0.11	$ & $	45.22	\pm	1.25	$ & $	44.97	\pm	1.46	$	\\
130702A	&	0.145	&	 	& $	1.05	\pm	0.44	$ & $	-10.40	\pm	0.05	$ & $	5.25	\pm	0.07	$ & $	39.34	\pm	1.54	$ & $	39.22	\pm	1.58	$	\\
130831A	&	0.4791	&	 	& $	1.91	\pm	0.03	$ & $	-9.40	\pm	0.04	$ & $	3.36	\pm	0.08	$ & $	43.56	\pm	1.19	$ & $	43.34	\pm	1.37	$	\\
130907A	&	1.238	&	4	& $	2.95	\pm	0.01	$ & $	-7.83	\pm	0.01	$ & $	2.81	\pm	0.02	$ & $	43.41	\pm	1.30	$ & $	43.08	\pm	1.57	$	\\
130925A	&	0.347	&	 	& $	1.49	\pm	0.02	$ & $	-10.29	\pm	0.03	$ & $	5.10	\pm	0.05	$ & $	40.49	\pm	1.15	$ & $	40.32	\pm	1.30	$	\\
131030A	&	1.295	&	4	& $	2.61	\pm	0.02	$ & $	-9.02	\pm	0.03	$ & $	3.08	\pm	0.06	$ & $	44.92	\pm	1.26	$ & $	44.63	\pm	1.50	$	\\
131103A	&	0.599	&	2	& $	2.01	\pm	0.24	$ & $	-9.68	\pm	0.09	$ & $	3.43	\pm	0.13	$ & $	44.32	\pm	1.36	$ & $	44.09	\pm	1.52	$	\\
131105A	&	1.686	&	6	& $	2.74	\pm	0.07	$ & $	-10.26	\pm	0.06	$ & $	3.93	\pm	0.16	$ & $	46.19	\pm	1.34	$ & $	45.89	\pm	1.58	$	\\
131117A	&	4.042	&	 	& $	2.35	\pm	0.07	$ & $	-10.78	\pm	0.08	$ & $	3.02	\pm	0.15	$ & $	48.88	\pm	1.30	$ & $	48.61	\pm	1.51	$	\\
131229A	&	1.04	&	 	& $	2.89	\pm	0.02	$ & $	-8.27	\pm	0.09	$ & $	2.47	\pm	0.13	$ & $	45.22	\pm	1.35	$ & $	44.90	\pm	1.61	$	\\
140206A	&	2.73	&	 	& $	2.65	\pm	0.02	$ & $	-9.25	\pm	0.01	$ & $	3.09	\pm	0.03	$ & $	45.57	\pm	1.25	$ & $	45.28	\pm	1.50	$	\\
140213A	&	1.2076	&	4	& $	2.25	\pm	0.01	$ & $	-9.60	\pm	0.07	$ & $	3.83	\pm	0.10	$ & $	43.70	\pm	1.23	$ & $	43.44	\pm	1.45	$	\\
140304A	&	5.283	&	 	& $	2.89	\pm	0.11	$ & $	-9.06	\pm	0.04	$ & $	2.35	\pm	0.06	$ & $	47.48	\pm	1.33	$ & $	47.16	\pm	1.59	$	\\
140419A	&	3.956	&	 	& $	3.16	\pm	0.13	$ & $	-9.51	\pm	0.04	$ & $	2.99	\pm	0.05	$ & $	47.66	\pm	1.38	$ & $	47.31	\pm	1.65	$	\\
140423A	&	3.26	&	 	& $	2.69	\pm	0.01	$ & $	-10.09	\pm	0.11	$ & $	3.33	\pm	0.16	$ & $	47.19	\pm	1.35	$ & $	46.89	\pm	1.59	$	\\
140506A	&	0.889	&	 	& $	2.57	\pm	0.03	$ & $	-9.76	\pm	0.07	$ & $	4.35	\pm	0.10	$ & $	43.54	\pm	1.28	$ & $	43.25	\pm	1.52	$	\\
140512A	&	0.725	&	 	& $	3.08	\pm	0.02	$ & $	-8.99	\pm	0.01	$ & $	3.44	\pm	0.03	$ & $	45.01	\pm	1.32	$ & $	44.67	\pm	1.60	$	\\
140518A	&	4.707	&	 	& $	2.40	\pm	0.08	$ & $	-10.37	\pm	0.07	$ & $	2.57	\pm	0.23	$ & $	49.11	\pm	1.37	$ & $	48.84	\pm	1.58	$	\\
140629A	&	2.275	&	8	& $	2.45	\pm	0.09	$ & $	-9.55	\pm	0.03	$ & $	2.73	\pm	0.06	$ & $	46.76	\pm	1.25	$ & $	46.49	\pm	1.48	$	\\
140703A	&	3.14	&	 	& $	2.94	\pm	0.07	$ & $	-9.98	\pm	0.05	$ & $	3.32	\pm	0.10	$ & $	47.50	\pm	1.34	$ & $	47.18	\pm	1.60	$	\\
140713A	&	0.935	&	 	& $	1.85	\pm	0.16	$ & $	-9.43	\pm	0.13	$ & $	3.13	\pm	0.28	$ & $	44.08	\pm	1.44	$ & $	43.87	\pm	1.59	$	\\
140907A	&	1.21	&	4	& $	2.49	\pm	0.01	$ & $	-9.42	\pm	0.06	$ & $	3.03	\pm	0.10	$ & $	45.79	\pm	1.26	$ & $	45.51	\pm	1.50	$	\\
141004A	&	0.574	&	 	& $	1.64	\pm	0.11	$ & $	-9.55	\pm	0.09	$ & $	2.86	\pm	0.15	$ & $	44.57	\pm	1.25	$ & $	44.38	\pm	1.40	$	\\
141121A	&	1.47	&	5	& $	2.29	\pm	0.21	$ & $	-11.63	\pm	0.06	$ & $	4.94	\pm	0.26	$ & $	46.06	\pm	1.46	$ & $	45.80	\pm	1.64	$	\\
141220A	&	1.3195	&	 	& $	2.62	\pm	0.01	$ & $	-8.84	\pm	0.12	$ & $	2.05	\pm	0.18	$ & $	47.08	\pm	1.36	$ & $	46.79	\pm	1.59	$	\\
141221A	&	1.452	&	5	& $	2.35	\pm	0.06	$ & $	-9.40	\pm	0.09	$ & $	2.83	\pm	0.14	$ & $	45.95	\pm	1.29	$ & $	45.68	\pm	1.50	$	\\
150206A	&	2.087	&	7	& $	2.85	\pm	0.07	$ & $	-9.66	\pm	0.24	$ & $	3.59	\pm	0.25	$ & $	45.81	\pm	1.55	$ & $	45.49	\pm	1.78	$	\\
150301B	&	1.5169	&	 	& $	2.66	\pm	0.03	$ & $	-9.72	\pm	0.10	$ & $	2.64	\pm	0.23	$ & $	47.90	\pm	1.40	$ & $	47.61	\pm	1.63	$	\\
150314A	&	1.758	&	6	& $	2.99	\pm	0.01	$ & $	-8.38	\pm	0.04	$ & $	2.43	\pm	0.05	$ & $	45.85	\pm	1.31	$ & $	45.52	\pm	1.59	$	\\
150323A	&	0.593	&	2	& $	2.18	\pm	0.07	$ & $	-10.73	\pm	0.10	$ & $	4.11	\pm	0.25	$ & $	45.66	\pm	1.37	$ & $	45.41	\pm	1.56	$	\\
150403A	&	2.06	&	7	& $	3.39	\pm	0.03	$ & $	-8.32	\pm	0.01	$ & $	2.84	\pm	0.01	$ & $	45.59	\pm	1.38	$ & $	45.22	\pm	1.68	$	\\
150821A	&	0.755	&	 	& $	2.69	\pm	0.03	$ & $	-10.23	\pm	0.18	$ & $	3.98	\pm	0.34	$ & $	45.90	\pm	1.58	$ & $	45.60	\pm	1.79	$	\\
151021A	&	2.33	&	8	& $	2.75	\pm	0.05	$ & $	-9.24	\pm	0.10	$ & $	2.85	\pm	0.11	$ & $	46.39	\pm	1.32	$ & $	46.08	\pm	1.57	$	\\
151027A	&	0.81	&	 	& $	2.56	\pm	0.03	$ & $	-9.12	\pm	0.01	$ & $	3.54	\pm	0.02	$ & $	43.93	\pm	1.24	$ & $	43.65	\pm	1.48	$	\\
151111A	&	3.5	&	11	& $	2.73	\pm	0.04	$ & $	-10.17	\pm	0.23	$ & $	2.75	\pm	0.38	$ & $	48.91	\pm	1.68	$ & $	48.60	\pm	1.88	$	\\
151229A	&	1.4	&	 	& $	2.40	\pm	0.06	$ & $	-9.26	\pm	0.03	$ & $	2.91	\pm	0.07	$ & $	45.46	\pm	1.24	$ & $	45.19	\pm	1.47	$	\\
160227A	&	2.38	&	 	& $	2.35	\pm	0.11	$ & $	-10.81	\pm	0.04	$ & $	4.31	\pm	0.09	$ & $	45.73	\pm	1.26	$ & $	45.46	\pm	1.48	$	\\
160509A	&	1.17	&	 	& $	2.89	\pm	0.01	$ & $	-9.86	\pm	0.04	$ & $	4.35	\pm	0.06	$ & $	44.50	\pm	1.30	$ & $	44.18	\pm	1.57	$	\\
160804A	&	0.736	&	 	& $	2.09	\pm	0.01	$ & $	-11.02	\pm	0.06	$ & $	5.29	\pm	0.18	$ & $	43.22	\pm	1.27	$ & $	42.98	\pm	1.47	$	\\
161001A	&	0.67	&	 	& $	2.79	\pm	0.07	$ & $	-8.75	\pm	0.05	$ & $	2.78	\pm	0.09	$ & $	45.43	\pm	1.31	$ & $	45.12	\pm	1.56	$	\\
161014A	&	2.823	&	 	& $	2.81	\pm	0.01	$ & $	-9.31	\pm	0.02	$ & $	2.58	\pm	0.06	$ & $	47.36	\pm	1.29	$ & $	47.05	\pm	1.55	$	\\
161017A	&	2.013	&	7	& $	2.86	\pm	0.02	$ & $	-10.22	\pm	0.05	$ & $	3.59	\pm	0.08	$ & $	47.22	\pm	1.30	$ & $	46.90	\pm	1.57	$	\\
161108A	&	1.159	&	3	& $	2.15	\pm	0.06	$ & $	-11.45	\pm	0.05	$ & $	5.11	\pm	0.25	$ & $	44.87	\pm	1.35	$ & $	44.63	\pm	1.54	$	\\
161117A	&	1.549	&	 	& $	2.31	\pm	0.01	$ & $	-10.12	\pm	0.05	$ & $	3.83	\pm	0.08	$ & $	45.14	\pm	1.23	$ & $	44.88	\pm	1.45	$	\\
161219B	&	0.1475	&	 	& $	2.02	\pm	0.10	$ & $	-9.63	\pm	0.02	$ & $	5.17	\pm	0.03	$ & $	39.87	\pm	1.20	$ & $	39.64	\pm	1.40	$	\\
170113A	&	1.968	&	 	& $	2.52	\pm	0.08	$ & $	-9.71	\pm	0.03	$ & $	3.25	\pm	0.05	$ & $	46.03	\pm	1.25	$ & $	45.75	\pm	1.49	$	\\
170202A	&	3.645	&	11	& $	3.06	\pm	0.24	$ & $	-10.11	\pm	0.04	$ & $	3.30	\pm	0.11	$ & $	48.14	\pm	1.46	$ & $	47.80	\pm	1.71	$	\\
170405A	&	3.51	&	11	& $	3.15	\pm	0.01	$ & $	-9.45	\pm	0.03	$ & $	2.63	\pm	0.07	$ & $	48.41	\pm	1.35	$ & $	48.06	\pm	1.63	$	\\
170604A	&	1.329	&	 	& $	2.71	\pm	0.12	$ & $	-9.96	\pm	0.09	$ & $	3.81	\pm	0.14	$ & $	45.71	\pm	1.35	$ & $	45.41	\pm	1.59	$	\\
170607A	&	0.557	&	1	& $	2.24	\pm	0.02	$ & $	-10.26	\pm	0.02	$ & $	4.72	\pm	0.05	$ & $	43.09	\pm	1.21	$ & $	42.83	\pm	1.42	$	\\
170705A	&	2.01	&	7	& $	2.47	\pm	0.03	$ & $	-9.92	\pm	0.02	$ & $	3.91	\pm	0.04	$ & $	44.79	\pm	1.23	$ & $	44.51	\pm	1.47	$	\\
170728B	&	1.272	&	4	& $	2.49	\pm	0.09	$ & $	-8.87	\pm	0.03	$ & $	3.02	\pm	0.06	$ & $	44.46	\pm	1.26	$ & $	44.18	\pm	1.49	$	\\
170903A	&	0.886	&	 	& $	2.25	\pm	0.03	$ & $	-10.69	\pm	0.05	$ & $	4.75	\pm	0.13	$ & $	44.12	\pm	1.25	$ & $	43.86	\pm	1.46	$	\\
171010A	&	0.3285	&	 	& $	2.26	\pm	0.00	$ & $	-10.16	\pm	0.09	$ & $	4.57	\pm	0.14	$ & $	43.26	\pm	1.27	$ & $	43.01	\pm	1.48	$	\\
171205A	&	0.0368	&	 	& $	2.10	\pm	0.39	$ & $	-11.82	\pm	0.05	$ & $	7.90	\pm	0.17	$ & $	38.71	\pm	1.54	$ & $	38.47	\pm	1.69	$	\\
171222A	&	2.409	&	9	& $	1.78	\pm	0.03	$ & $	-11.68	\pm	0.07	$ & $	4.69	\pm	0.35	$ & $	45.64	\pm	1.46	$ & $	45.43	\pm	1.60	$	\\
180205A	&	1.409	&	 	& $	1.93	\pm	0.09	$ & $	-9.91	\pm	0.07	$ & $	3.23	\pm	0.11	$ & $	45.23	\pm	1.23	$ & $	45.01	\pm	1.41	$	\\
180314A	&	1.445	&	5	& $	2.40	\pm	0.01	$ & $	-9.98	\pm	0.06	$ & $	3.36	\pm	0.12	$ & $	46.16	\pm	1.26	$ & $	45.89	\pm	1.49	$	\\
180325A	&	2.248	&	8	& $	3.00	\pm	0.06	$ & $	-8.93	\pm	0.01	$ & $	2.68	\pm	0.04	$ & $	46.62	\pm	1.32	$ & $	46.29	\pm	1.59	$	\\
180329B	&	1.998	&	7	& $	2.16	\pm	0.08	$ & $	-10.54	\pm	0.04	$ & $	3.49	\pm	0.09	$ & $	46.67	\pm	1.23	$ & $	46.43	\pm	1.44	$	\\
180620B	&	1.1175	&	3	& $	2.57	\pm	0.06	$ & $	-10.05	\pm	0.05	$ & $	4.04	\pm	0.08	$ & $	45.02	\pm	1.27	$ & $	44.73	\pm	1.51	$	\\
180720B	&	0.654	&	2	& $	3.02	\pm	0.01	$ & $	-8.35	\pm	0.01	$ & $	3.52	\pm	0.02	$ & $	43.11	\pm	1.31	$ & $	42.77	\pm	1.59	$	\\
180728A	&	0.117	&	 	& $	1.94	\pm	0.01	$ & $	-8.74	\pm	0.01	$ & $	3.95	\pm	0.02	$ & $	40.53	\pm	1.17	$ & $	40.31	\pm	1.36	$	\\
181010A	&	1.39	&	 	& $	2.15	\pm	0.35	$ & $	-9.42	\pm	0.04	$ & $	3.05	\pm	0.07	$ & $	44.96	\pm	1.45	$ & $	44.72	\pm	1.61	$	\\
181020A	&	2.938	&	10	& $	3.19	\pm	0.01	$ & $	-8.29	\pm	0.01	$ & $	2.03	\pm	0.03	$ & $	47.07	\pm	1.34	$ & $	46.72	\pm	1.63	$	\\
181110A	&	1.505	&	5	& $	2.12	\pm	0.38	$ & $	-10.15	\pm	0.13	$ & $	3.36	\pm	0.16	$ & $	45.95	\pm	1.56	$ & $	45.72	\pm	1.70	$	\\
190106A	&	1.859	&	 	& $	2.69	\pm	0.18	$ & $	-9.88	\pm	0.02	$ & $	3.85	\pm	0.05	$ & $	45.33	\pm	1.33	$ & $	45.03	\pm	1.56	$	\\
190114C	&	0.425	&	 	& $	3.17	\pm	0.01	$ & $	-7.39	\pm	0.03	$ & $	2.75	\pm	0.04	$ & $	42.97	\pm	1.34	$ & $	42.62	\pm	1.63	$	\\
190324A	&	1.1715	&	 	& $	2.46	\pm	0.01	$ & $	-9.03	\pm	0.06	$ & $	3.01	\pm	0.06	$ & $	44.79	\pm	1.24	$ & $	44.52	\pm	1.47	$	\\
190719C	&	2.469	&	9	& $	2.47	\pm	0.03	$ & $	-10.64	\pm	0.06	$ & $	4.29	\pm	0.13	$ & $	45.66	\pm	1.28	$ & $	45.38	\pm	1.51	$	\\
190829A	&	0.0785	&	 	& $	1.07	\pm	0.04	$ & $	-9.10	\pm	0.02	$ & $	4.61	\pm	0.03	$ & $	37.77	\pm	1.14	$ & $	37.64	\pm	1.24	$	\\
191004B	&	3.503	&	11	& $	2.89	\pm	0.06	$ & $	-9.13	\pm	0.21	$ & $	2.34	\pm	0.41	$ & $	47.69	\pm	1.72	$ & $	47.37	\pm	1.93	$	\\
191011A	&	1.722	&	6	& $	2.18	\pm	0.15	$ & $	-10.18	\pm	0.08	$ & $	2.65	\pm	0.16	$ & $	47.92	\pm	1.32	$ & $	47.67	\pm	1.51	$	\\
191221B	&	1.148	&	3	& $	2.91	\pm	0.03	$ & $	-8.62	\pm	0.02	$ & $	2.81	\pm	0.04	$ & $	45.29	\pm	1.30	$ & $	44.97	\pm	1.57	$	\\
200205B	&	1.465	&	5	& $	2.33	\pm	0.08	$ & $	-10.93	\pm	0.09	$ & $	3.91	\pm	0.28	$ & $	46.98	\pm	1.42	$ & $	46.72	\pm	1.62	$	\\
200829A	&	1.25	&	4	& $	2.86	\pm	0.00	$ & $	-8.05	\pm	0.01	$ & $	2.64	\pm	0.01	$ & $	44.17	\pm	1.28	$ & $	43.86	\pm	1.55	$	\\
201020A	&	2.903	&	10	& $	2.12	\pm	0.06	$ & $	-10.59	\pm	0.05	$ & $	3.42	\pm	0.13	$ & $	46.88	\pm	1.24	$ & $	46.64	\pm	1.44	$	\\
201021C	&	1.07	&	3	& $	2.41	\pm	0.16	$ & $	-9.40	\pm	0.13	$ & $	2.61	\pm	0.27	$ & $	46.62	\pm	1.48	$ & $	46.35	\pm	1.68	$	\\
210104A	&	0.46	&	 	& $	2.45	\pm	0.06	$ & $	-8.82	\pm	0.01	$ & $	3.18	\pm	0.02	$ & $	43.81	\pm	1.23	$ & $	43.54	\pm	1.47	$	\\
210210A	&	0.715	&	 	& $	1.45	\pm	0.26	$ & $	-9.43	\pm	0.02	$ & $	3.38	\pm	0.09	$ & $	42.56	\pm	1.31	$ & $	42.39	\pm	1.43	$	\\
210411C	&	2.826	&	 	& $	1.75	\pm	0.40	$ & $	-9.45	\pm	0.05	$ & $	2.73	\pm	0.12	$ & $	44.92	\pm	1.52	$ & $	44.72	\pm	1.63	$	\\
210610A	&	3.54	&	11	& $	2.50	\pm	0.16	$ & $	-10.32	\pm	0.04	$ & $	3.07	\pm	0.11	$ & $	47.97	\pm	1.32	$ & $	47.69	\pm	1.54	$	\\
210610B	&	1.13	&	3	& $	2.82	\pm	0.01	$ & $	-8.88	\pm	0.03	$ & $	3.06	\pm	0.05	$ & $	45.12	\pm	1.28	$ & $	44.81	\pm	1.55	$	\\
210619B	&	1.937	&	 	& $	2.79	\pm	0.01	$ & $	-7.90	\pm	0.02	$ & $	2.56	\pm	0.03	$ & $	43.85	\pm	1.27	$ & $	43.54	\pm	1.53	$	\\
210702A	&	1.16	&	3	& $	2.91	\pm	0.07	$ & $	-8.21	\pm	0.02	$ & $	2.77	\pm	0.03	$ & $	44.37	\pm	1.30	$ & $	44.04	\pm	1.57	$	\\
210722A	&	1.145	&	3	& $	2.38	\pm	0.12	$ & $	-9.57	\pm	0.05	$ & $	3.23	\pm	0.10	$ & $	45.41	\pm	1.28	$ & $	45.14	\pm	1.49	$	\\
210731A	&	1.2525	&	4	& $	2.64	\pm	0.04	$ & $	-11.05	\pm	0.14	$ & $	4.24	\pm	0.40	$ & $	47.19	\pm	1.64	$ & $	46.89	\pm	1.84	$	\\
210822A	&	1.736	&	6	& $	3.04	\pm	0.03	$ & $	-7.45	\pm	0.02	$ & $	1.95	\pm	0.02	$ & $	44.82	\pm	1.32	$ & $	44.49	\pm	1.59	$	\\
210905A	&	6.318	&	 	& $	3.17	\pm	0.10	$ & $	-10.96	\pm	0.05	$ & $	3.96	\pm	0.10	$ & $	48.90	\pm	1.38	$ & $	48.55	\pm	1.66	$	\\
220101A	&	4.618	&	 	& $	3.21	\pm	0.03	$ & $	-9.48	\pm	0.06	$ & $	3.24	\pm	0.07	$ & $	47.05	\pm	1.37	$ & $	46.69	\pm	1.65	$	\\
220117A	&	4.961	&	 	& $	2.63	\pm	0.09	$ & $	-10.40	\pm	0.04	$ & $	2.93	\pm	0.10	$ & $	48.80	\pm	1.29	$ & $	48.51	\pm	1.53	$	\\
220521A	&	5.6	&	 	& $	2.53	\pm	0.09	$ & $	-9.92	\pm	0.06	$ & $	2.38	\pm	0.15	$ & $	48.76	\pm	1.31	$ & $	48.48	\pm	1.54	$	\\
221009A	&	0.1505	&	 	& $	2.57	\pm	0.10	$ & $	-7.89	\pm	0.04	$ & $	4.26	\pm	0.04	$ & $	39.08	\pm	1.27	$ & $	38.79	\pm	1.51	$	\\
230818A	&	2.42	&	9	& $	2.86	\pm	0.05	$ & $	-10.08	\pm	0.06	$ & $	3.40	\pm	0.19	$ & $	47.38	\pm	1.38	$ & $	47.06	\pm	1.63	$	\\
231118A	&	0.8304	&	 	& $	2.60	\pm	0.05	$ & $	-9.31	\pm	0.06	$ & $	3.37	\pm	0.14	$ & $	44.91	\pm	1.30	$ & $	44.62	\pm	1.54	$	\\
231210B	&	3.13	&	 	& $	2.86	\pm	0.08	$ & $	-9.60	\pm	0.12	$ & $	2.58	\pm	0.19	$ & $	48.20	\pm	1.41	$ & $	47.88	\pm	1.66	$	\\
\hline
\end{longtable}
\tablefoot{GRBs belonging to the calibration sub-samples are labeled as in Table~\ref{tab:no3}. Errors are at 1-sigma confidence level. The distance moduli derived from \emph{flat} and \emph{curved} calibrations (see Table~\ref{tab:no5}) are labeled with $\Omega_k=0$ and $\Omega_k\neq0$, respectively.}
}

\section{Standardization of the C244 catalog with the ``max-model'' parametrization of SNe Ia}\label{appendix2}
\begin{figure*}[htb!]
\centering
\includegraphics[width=0.95\hsize,clip]{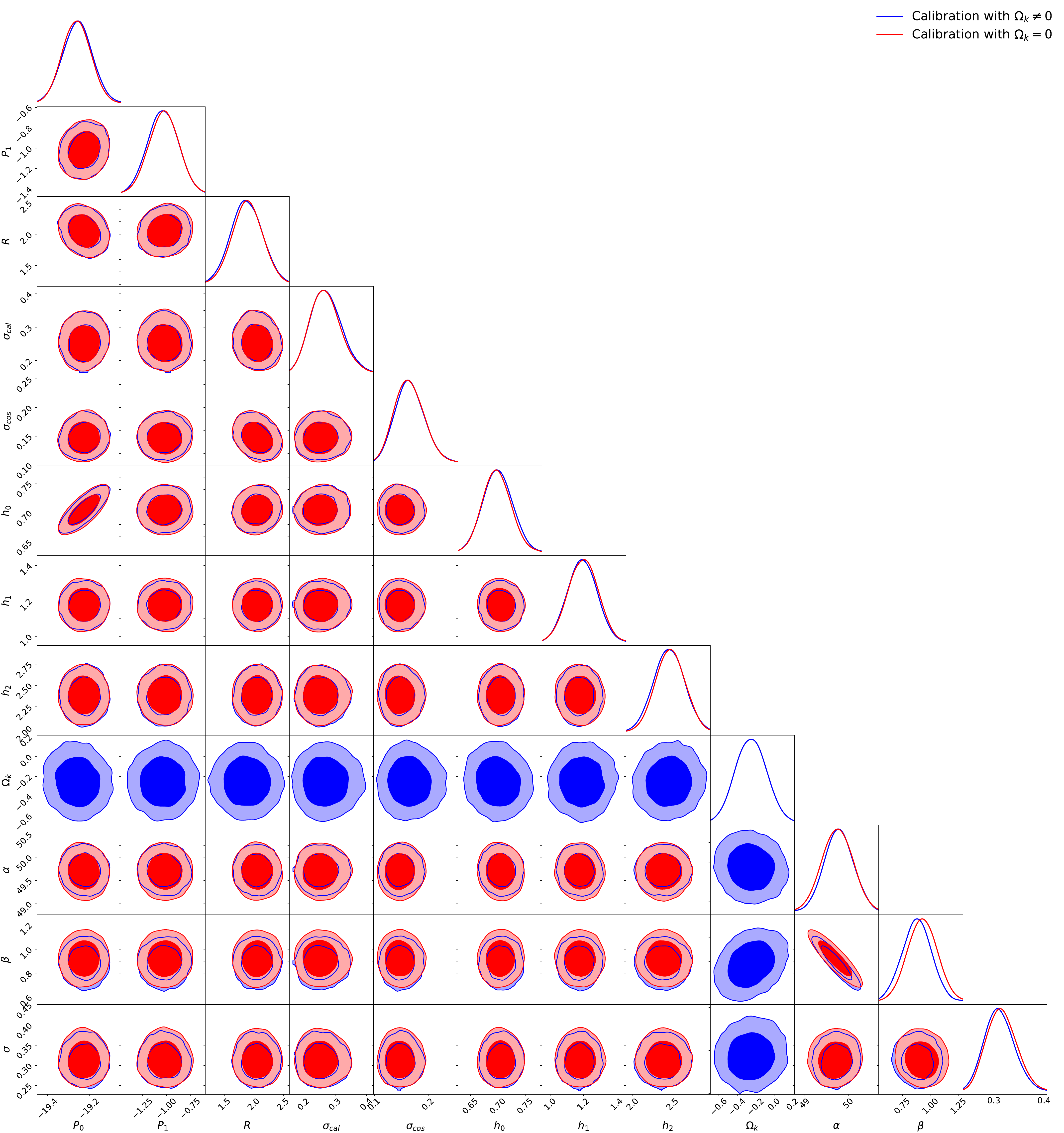}
\caption{MCMC posteriors on the parameters involved in the standardization of the C244 catalog, for the cases $\Omega_k=0$ (red) and $\Omega_k\neq0$ (blue), using the ``max-model'' parametrization of SNe Ia. The dark (light) shaded regions correspond to $1$-sigma ($2$-sigma) confidence levels. The best-fit values are listed in Table~\ref{tab:no5}. For each parameter one-dimensional LLHs are also shown.}
\label{fig:no6a}
\end{figure*}

\normalsize
The calibration constraints obtained from Eqs.~\eqref{bayes}--\eqref{priors} are displayed in Fig.~\ref{fig:no6a} for both cases $\Omega_k=0$ and $\Omega_k\neq0$.

\section{Standardization of the C244 catalog with the SALT2 parametrization of SNe Ia}\label{appendix3}
\normalsize
In the SALT2 modeling of SN Ia light curves, the uncorrected brightness $m_B$ is linked to the distance modulus $\mu$, the absolute magnitude $M$ of a fiducial SN Ia, the stretch parameter $x_1$ describing the light-curve width, and the color parameter $c$ through the following relation \citep[see, e.g.,][]{2022ApJ...938..113S}
\begin{equation}
\label{eq:SALT2}
m_B = M - \alpha_{\rm S} x_1 + \beta_{\rm S} c + \mu \,.
\end{equation}
The nuisance parameters $\alpha_{\rm S}$ and $\beta_{\rm S}$ encode the empirical correlations among luminosity, stretch, and color.

From the online Pantheon+ catalog, we selected the SNe and the corresponding SALT2 parameters of: (i) the IR--SBF calibrator sample from \citet{Garnavich:2022hef} and (ii) the SH0ES calibrator sample and the cosmological sample used in \citet{2021A&A...647A..72K}.
The distance moduli in Eq.~\eqref{eq:SALT2}
are taken from \citet{Jensen:2021ooi} for the IR--SBF calibrator sample, from the Cepheids for the SH0ES calibrator sample, and from Eq.~\eqref{eq:mu} for the cosmological sample.
\begin{figure*}[t]
\centering
\includegraphics[width=0.95\hsize,clip]{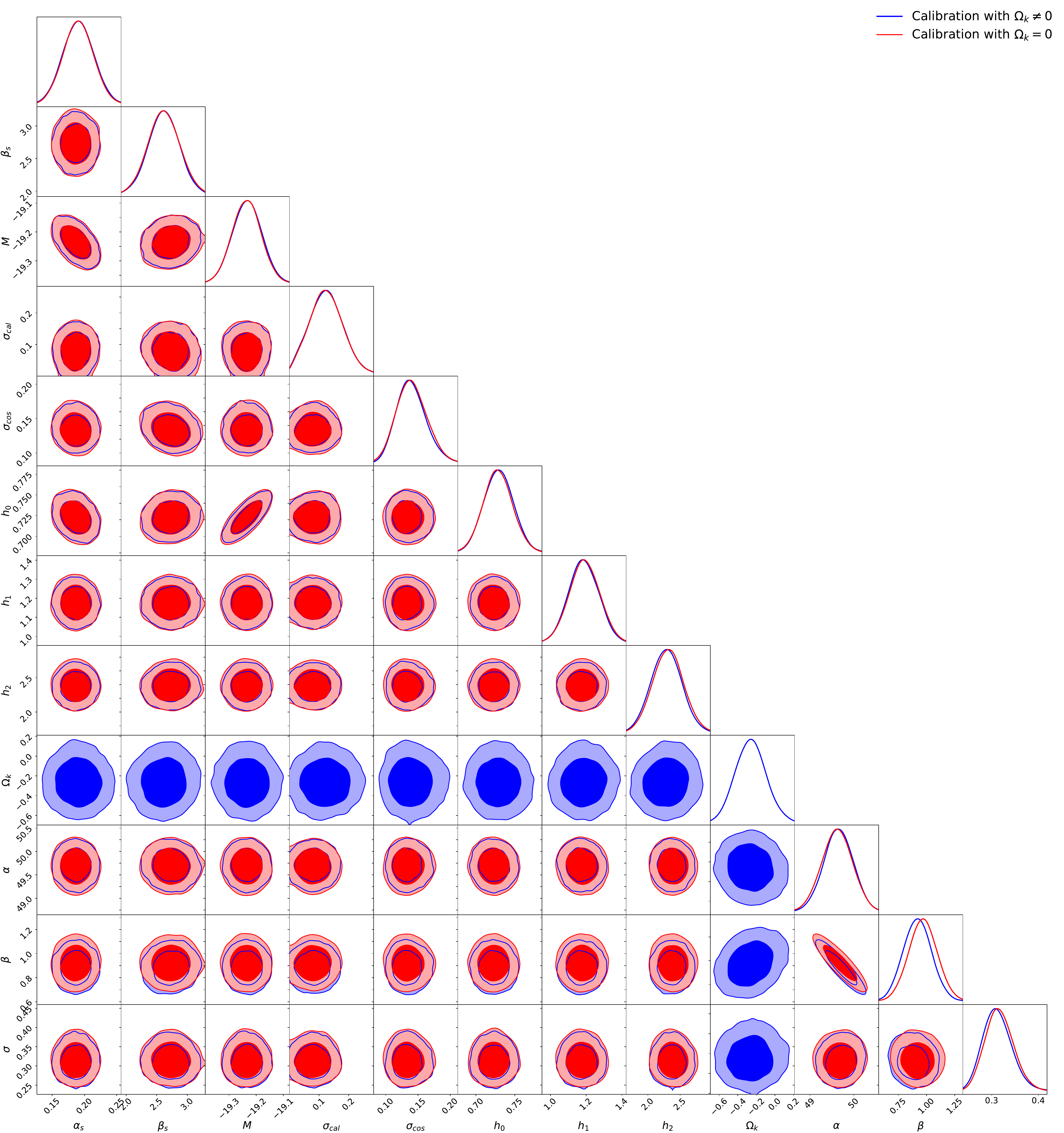}
\caption{MCMC posteriors on the parameters involved in the standardization of the C244 catalog, for the cases $\Omega_k=0$ (red) and $\Omega_k\neq0$ (blue), using the SALT2 parametrization of SNe Ia. The dark (light) shaded regions correspond to $1$-sigma ($2$-sigma) confidence levels. The best-fit values are listed in Table~\ref{tab:no5b}. For each parameter one-dimensional LLHs are also shown.}
\label{fig:no6b}
\end{figure*}

Since the Pantheon+ catalog contains multiple entries for the same SN from different surveys or repeated observations, to avoid duplication we selected the best-quality SN fit according to the following ordered criteria: (i) the highest SNANA fit probability (\texttt{FITPROB}); (ii) if the entries have comparable \texttt{FITPROB} values, the lowest reduced chi-square; and (iii) if a further tie remains, the entry with the smallest uncertainty on $m_B$.

The calibration constraints obtained from the SALT2 parametrization are displayed in Fig.~\ref{fig:no6b} for both cases $\Omega_k=0$ and $\Omega_k\neq0$.
Here, using Eq.~\eqref{eq:SALT2}, the LLH functions for the calibrator (IR--SBF~+~SH0ES) and the cosmological samples can be written as, respectively,
\begin{subequations}
\begin{align}
\ln \mathcal{L}_{cal} & = -\frac{1}{2}\sum_{i=1}^{N_{cal}} \left[ \frac{(m_B^i - m_B)^2}{\Sigma_{cal,i}^2} + \ln \left(2 \pi \Sigma_{cal,i}^2\right)\right]\,,\\ 
\ln \mathcal{L}_{cos} &= -\frac{1}{2}\sum_{j=1}^{N_{cos}} \left[ \frac{(m_B^j - m_B)^2}{\Sigma_{cos,j}^2} + \ln \left(2 \pi \Sigma_{cos,j}^2\right) \right]\,, 
\end{align}
\end{subequations}
with the corresponding variances
\begin{subequations}
\begin{align}
\Sigma_{cal,i}^2 &= \sigma_{m_{B,i}}^2 + \sigma_{\mu,i}^2 + \alpha_{\rm S}^2 \sigma_{x_{1,i}}^2 + \beta_{\rm S}^2 \sigma_{c,i}^2 + 2 \alpha_{\rm S} C_{m_B x_1,i} - 2 \beta_{\rm S} C_{m_B c,i} - 
 2 \alpha_{\rm S} \beta_{\rm S} C_{x_1 c,i} + \sigma_{cal}^2\,, \\
\Sigma_{cos,j}^2 &= \sigma_{m_{B,j}}^2 + \alpha_{\rm S}^2 \sigma_{x_{1,j}}^2 + \beta_{\rm S}^2 \sigma_{c,j}^2 + 2 \alpha_{\rm S} C_{m_B x_1,j} - 2 \beta_{\rm S} C_{m_B c,j} - 
 2 \alpha_{\rm S} \beta_{\rm S} C_{x_1 c,j} + \sigma_{cos}^2\,, 
\end{align}
\end{subequations}
where, as in Sec.~\ref{sec:3.2}, all the uncertainties and the extra dispersions are parametrized by $\sigma_{cal}$ (assumed to be the same for IR--SBF and SH0ES) and $\sigma_{cos}$. The symbols $C_{yz,i}$ are the covariance terms between the observables $y$ and $z$.

\end{appendix}

\end{document}